# Discovery of a Topological Charge Density Wave

**Authors:** Maksim Litskevich[1]*, Md Shafayat Hossain[1]*†, Songbo Zhang[2]*, Zi-Jia Cheng[1], Satya N. Guin[3], Nitesh Kumar[3], Chandra Shekhar[3], Zhiwei Wang[4,5], Yongkai Li[4,5], Guoqing Chang[6], Jia-Xin Yin[1], Qi Zhang[1], Guangming Cheng[7], Tyler A. Cochran[1], Nana Shumiya[1], Yu-Xiao Jiang[1], Xian P. Yang[1], Daniel Multer[1], Xiaoxiong Liu[2], Nan Yao[7], Yugui Yao[4,5], Claudia Felser[3], Titus Neupert[2], M. Zahid Hasan[1,8]†

**Affiliations:**
[1]Laboratory for Topological Quantum Matter and Advanced Spectroscopy (B7), Department of Physics, Princeton University, Princeton, New Jersey, USA.

[3]Department of Physics, University of Zurich, Winterthurerstrasse, Zurich, Switzerland.

[2]Max Planck Institute for Chemical Physics of Solids, 16 Nöthnitzer Straße 40, 01187 Dresden, Germany.

[4]Centre for Quantum Physics, Key Laboratory of Advanced Optoelectronic Quantum Architecture and Measurement (MOE), School of Physics, Beijing Institute of Technology, Beijing 100081, China.

[5]Beijing National Laboratory for Condensed Matter Physics and Institute of Physics, Chinese Academy of Sciences, Beijing 100190, China.

[6]Division of Physics and Applied Physics, School of Physical and Mathematical Sciences, Nanyang Technological University, Singapore.

[7]Princeton Institute for Science and Technology of Materials, Princeton University, Princeton, NJ, USA.

[8]Lawrence Berkeley National Laboratory, Berkeley, California 94720, USA.

†Corresponding authors, E-mail: mdsh@princeton.edu; mzhasan@princeton.edu.

*These authors contributed equally to this work.

## Abstract
**Charge density waves appear in numerous condensed matter platforms, ranging from high-temperature superconductors to quantum Hall systems. Despite such ubiquity, there has been a lack of direct experimental study on boundary states that can uniquely stem from the charge order. Here, using scanning tunneling microscopy, we directly visualize the bulk and boundary phenomenology of the charge density wave in a topological material, $Ta_2Se_8I$. Below the charge density wave transition temperature ($T_{CDW} \simeq 260$ K), tunneling spectra on an atomically resolved lattice reveal a large insulating gap in the bulk and on the surface, exceeding 500 meV, surpassing predictions from standard weakly-coupled mean-field theory. Spectroscopic imaging confirms the presence of a charge density wave, with local density of states maxima at the conduction band corresponding to the local density of states minima at the valence band, thus revealing a π phase difference in the respective charge density wave order. Concomitantly, at a monolayer step edge, we detect an in-gap boundary mode (edge state) with modulations along the edge that match the charge density wave wavevector along the edge. Intriguingly, the phase of the edge state modulation shifts by π within the charge order gap, connecting the fully gapped bulk (and surface) conduction and valence bands via a smooth energy-phase relation. This bears similarity to the topological spectral flow of edge modes, where the boundary modes bridge the gapped bulk modes in energy and momentum *magnitude* but in $Ta_2Se_8I$, the connectivity distinctly occurs in energy and momentum *phase*. Notably, our temperature-dependent measurements indicate a vanishing of the insulating gap and the in-gap edge state above $T_{CDW}$,**

**suggesting their direct relation to the charge order. The theoretical analysis also indicates that the observed boundary mode is likely topological and linked to the charge order. Our study further illuminates a puzzle concerning the enigmatic ground state of $Ta_2Se_8I$. It is theoretically envisioned to be the heavily sought-after axion insulator phase featuring a topological surface state that may arise when a topological semimetal turns insulating through the charge order. Contrary to this theoretical expectation, our experimental bulk-surface-step edge phenomenology reveals the absence of a surface state and the presence of an edge state instead. This unexpected observation challenges the existing axion insulator interpretation of the charge ordered phase, while at the same time suggests that it carries topological properties.**

## Main text

Axion particles, originally proposed as an explanation for the absence of charge–parity violation in the strong interaction between quarks and as a candidate for dark matter, found their manifestation in condensed matter platforms such as topological antiferromagnets, ferromagnetic topological insulators, charge density wave Weyl semimetals[1-10]. $Ta_2Se_8I$ is the only known material platform to date[11-14] where, in theory, a charge density wave transition from a Weyl semimetal state could lead to an axion insulator phase. Here the charge density wave, whose sliding mode (phason) is an axion, supposedly connects the pair of Weyl points with opposite chirality at different positions in the momentum space, thus breaking the chiral symmetry of the Weyl semimetal, opening an insulating gap (Fig. 1**a**), and tentatively realizing an axion insulator. An axion insulator features (*i*) a quantized, bulk magnetoelectric coupling coefficient of $\pi$[15-17] that is challenging to access experimentally and (*ii*) a gapless, time-reversal symmetry protected topological surface state[12] that is readily accessible in spectroscopic techniques. Still, the topology of the charge density wave insulator phase of $Ta_2Se_8I$ remains unknown and subject to debate[4,5,11-14,18]. Here employing scanning tunneling microscopy, we uncover a boundary mode (edge state) residing within the bulk charge density wave gap and unveil an intimate connection between the edge mode and the charge ordered state. Notably, we also find that there is no detectable surface state present within the charge density wave[19]. This is incompatible with the prevailing assumption of an axion insulator ground state in $Ta_2Se_8I$, which must carry a surface Dirac cone in a nonmagnetic material. Taken together, these observations hint at a unique nature of the charge density wave state, different from an axion insulator.

$Ta_2Se_8I$ has a quasi-one-dimensional crystal structure with Ta and I atomic chains running along the *c*-axis of the crystal (Fig. 1**b**). Ta atoms form strong covalent bonds with adjacent Se atoms whereas I atoms are weakly tied to Ta-Se chiral chains. $Ta_2Se_8I$ crystals grow in a few mm-long needle shape with an aspect ratio of approximately 10, reflecting its one-dimensional nature (see inset of Fig. 1**c**). Scanning transmission electron microscopy perpendicular to the growth direction (*c*) reveals a squared arrangement of Ta and I atoms (Fig. 1**c**). Due to a weak interaction of I atoms with Ta-Se chains, $Ta_2Se_8I$ has a natural cleaving plane (110) as marked in Fig. 1**b**. Using scanning transmission electron microscopy and electron diffraction analysis, we confirm that the cleaving plane of our samples is (110); see Extended Fig. 1 for details. We cleave $Ta_2Se_8I$ samples in-situ in ultra-high vacuum conditions ($< 5 \times 10^{-10}$ mbar) for our scanning tunneling microscopy measurements. In the scanning tunneling microscopy image on a clean, freshly cleaved sample, we find that the surface contains I atoms, 50% of which are expelled from the original honeycomb-like lattice (Fig. 1**d**), leaving a close-to-squared I lattice with the following interatomic distances (marked in Fig. 1**d** right panel): $d \simeq 9.5$ Å, $c \simeq 13$ Å, $\Theta \simeq 86\pm0.5°$ (Figs. 1**e**, **f**); this is consistent with the prior scanning tunneling microscopy data[20,21]. Notably, the topography taken at $T = 160$ K visualizes a one-dimensional charge density wave (Fig. 1**e**). The Fourier transform of the large-scale drift corrected topography[22] image reveals Bragg peaks with a close-to-squared lattice symmetry, as well as the charge density wave wavevector peaks (Fig. 1g). We determine the charge density wave period, $\lambda_{CDW}$, to be $9.1 \pm 0.6$ nm. The charge density wave wavevector, $\vec{q}_{CDW} \simeq (0.054\frac{2\pi}{a}, -0.054\frac{2\pi}{a}, 0.098\frac{2\pi}{c})$, lies in the (110) plane; see Methods and Extended Figs. 2-3 for details. While the charge density wave wavelength remains relatively uniform within a single batch of samples, it shows slight variations across samples from different batches, averaging at $<\lambda_{CDW}>$ = $10.4 \pm 1.6$ nm. Our result is consistent with X-ray and neutron scattering experiments[23-29] but differs slightly from a recent scanning tunneling microscopy data which reports an enlarged $\lambda_{CDW} = 17 \pm 1$ nm attributed to the growth condition and doping effect of abundant I-vacancies[20,21].

Having explored the crystal structure of $Ta_2Se_8I$, we focus on tunneling spectroscopy measurements. First, we show two d$I$/d$V$ spectroscopic maps, taken at the same location at -0.60 V and 0.45 V, in Fig. 1h. The d$I$/d$V$ maps crucially reveal that the local density of states maxima at 0.45 V corresponds to the local density of states minima at -0.60 V, signifying a spectroscopic contrast reversal upon adjusting the tip-sample bias from the conduction band to the valence band (Fig. 1k). This contrast switch is typically associated with the electronic character of the charge order[30-32]. The Fourier transform images of the spectroscopic maps in Fig. 1h, illustrated in Fig. 1i, unambiguously reveal the presence of long-wavelength charge density wave peaks that align well with the corresponding peaks from topography. Next, we turn to energy-resolved d$I$/d$V$ spectroscopy. It is first worth emphasizing that, as shown in Fig. 1j (left panel), $\lambda_{CDW}$ remains constant across varying energy levels, a behavior anticipated from a charge density wave state. At $T$ = 160 K, spatially resolved differential spectra taken on the atomic, free-of-defect surface reveal a large insulating gap (Extended Fig. 4b); see Methods and Extended Fig. 5 for the details on energy gap determination. Analyzing the average d$I$/d$V$ spectrum, we find the energy gap, $\Delta_{CDW}$ to be $\simeq$ 550 meV (see Extended Fig. 4b and the right panel of Fig. 1j). It is worth comparing our $\Delta_{CDW}$ with the gaps obtained from other techniques. The energy difference between the Fermi level and valence band in our tunneling spectra is $\simeq$260 meV, which matches well with our photoemission spectroscopy data (Extended Fig. 6) where the energy difference between the Fermi energy and the emergence of spectral weight density below the Fermi level is $\simeq$250 meV. These values are consistent with the previous photoemission spectroscopy experiments[14,20,33]. On the other hand, our tunneling $\Delta_{CDW}$ appears to be larger than the gap observed in transport measurements[5,20] and recent scanning tunneling microscopy data[21]; it is possible that the crystal defects and numerous I-vacancies (that manifests through a larger $\lambda_{CDW}$) in their sample can cause the discrepancy between the results (see Methods). To investigate the origin of the large energy gap, we compare our experimental results to the outcomes of the mean-field theory of a charge density wave[34]. The mean-field formula that relates the magnitude of the charge density wave gap and the transition temperature assuming weak coupling goes as $\Delta_{CDW}/3.52k_B T_{CDW} \simeq 1$. Note that it has a direct analogy to Bardeen–Cooper–Schrieffer theory of superconductivity[34]. Here in $Ta_2Se_8I$, however, $\Delta_{CDW}/3.52k_B T_{CDW} \simeq 7$ ($\Delta_{CDW} \approx 550$ meV, $T_{CDW} \simeq 260$ K), is nearly an order of magnitude larger than what is expected from a traditional, weak coupling charge density wave. It indicates the presence of strong coupling in the charge density wave state of $Ta_2Se_8I$[35,36]. In the context of the large $\Delta_{CDW}$, it is worth noting that our observation of a substantial insulating energy gap in the surface, accompanied by the absence of discernible in-gap d$I$/d$V$ intensity, indicates the absence of surface states in the time-reversal symmetric material $Ta_2Se_8I$. Our photoemission spectroscopy data supports this absence (Extended Fig. 6). As discussed earlier, this observation contradicts the existence of the axion insulator state in $Ta_2Se_8I$. Furthermore, the energy scale of the spin-orbit coupling, which determines the energy window of the Weyl physics and the energy gap of an axion insulator, is approximately $\Delta_{SO} \simeq 20$ meV in $Ta_2Se_8I$[37]. This value is significantly smaller than the observed $\Delta_{CDW}$. The substantial difference in energy scales between the axion insulator and $\Delta_{CDW}$ strongly suggests that the axion insulator state is highly improbable in $Ta_2Se_8I$.

We further investigate the large charge density wave gap by examining its spatial dependence. While the magnitude of the tunneling spectra is dependent on both tip and sample, the energy gap usually correlates with the electronic nature of the sample only. In our case, spatially resolved tunneling spectra taken along the charge density wave vector direction reveal a spatially modulated energy gap with a modulation amplitude of $\simeq 40$ meV and periodicity of $\simeq 12 \pm 1$ nm (Extended Figs. 4**a-c**). This periodicity closely matches the charge density wave period of the specific sample, which has been extracted from topography. Moreover, comparing the topography and the spatial dependence of the charge density wave gap, we find that the topographic maxima (minima) correspond to the gap minima (maxima). Such a spatial gap modulation associated with the charge density wave has been seen in several charge density wave compounds[30,38] and interpreted as evidence for the electronic nature of the charge density wave. Taken collectively, our observations of direct correlations between *(i)* the charge density wave in topography, *(ii)* charge density wave wavevector peaks in the Fourier transform of the spectroscopic map, and *(iii)* spatially modulated charge density wave gap, alongside the demonstrated spectroscopic contrast reversal upon altering the tip-sample bias from the conduction band to the valence band, all consistently support the electronic nature for the charge density wave in $Ta_2Se_8I$.

Having discussed the electronic nature of the charge density wave in $Ta_2Se_8I$, we investigate the topological properties of the charge density wave state. Figure 2 captures the highlight of our experiment - the observation of an in-gap edge state within the charge density wave. Scanning tunneling microscopy topography image in Fig. 2**a** illustrates an atomically sharp monolayer step edge along the *c* direction; see Extended Fig. 1 for the analysis of edge orientations. The height line profile perpendicular to the edge clearly shows uniform atomic correlations on the two sides of the monolayer step edge (Fig. 2**b**). The phase of the charge density wave appears to be continuous across the edge, consistent with the bulk nature (not from trivial surface effects) of the charge density wave; see Methods and Extended Fig. 3 for more details. To probe the electronic states localized at the edge, we perform d$I$/d$V$ measurements along the line (Fig. 2**a**) passing through the edge. Away from the edge, the tunneling spectra feature a large insulating gap (blue curves in Fig. 2**c**). In sharp contrast, the spectra taken at the edge (orange curves in Fig. 2**c**) reveal a pronounced, finite density of states within the insulating gap. The presence of the in-gap state localized at the edge suggests the existence of edge modes[39-55] in the charge density wave phase of $Ta_2Se_8I$.

To further explore the edge state, in Fig. 2**d** we present d$I$/d$V$ spectroscopic maps taken at a monolayer step edge identified by a topographic image (Fig. 2**d** bottom panel). In the real-space differential conductance maps at the Fermi level and $E = \pm 100$ meV, we find that the monolayer step edge exhibits pronounced edge states manifested via increased differential conductance within the insulating gap. In contrast, the spectroscopic image at -0.8 V (Fig. 2**d** top panel) exhibits suppressed differential conductance along the edge. Figure 2**e**, which depicts the spatial spread of the edge state, reveals an exponential decay of the edge state on the crystal side of the step edge. The edge state decays with a characteristic length of $r_0 \sim 1.25$ nm on the crystal side, indicating a strongly confined nature of the edge state. Note that this is nearly 50 times smaller than in HgTe/CdTe quantum wells and 2 times larger than in Bismuthene[39,43,55]. The strongly confined nature of the edge state is evident in Extended Fig. 7, displaying line profiles for the topography and corresponding d$I$/d$V$ around the step edge. These profiles illustrate that the edge state exhibits exponential localization within the crystal side, diminishing as it extends towards the vacuum side.

With the electronic nature and the edge states associated with the charge density wave insulator phase in $Ta_2Se_8I$ now addressed, we shift our focus to its temperature dependence. In Fig. 3**a**, we show large-scale topography images and the corresponding Fourier transforms (inset) at 160 K, 250 K, and 280 K. Charge density wave and the corresponding Fourier transform peaks are detected at 160 K and 250 K topography images. On the other hand, neither the charge density wave nor its Fourier transform peaks are seen in 280 K data (Fig. 3**a** right panel). This observation is consistent with recent scanning tunneling microscopy data[21] and is corroborated by our transport measurement shown in Fig. 3**b**, which indicates that the charge density wave transition occurs at $T_{CDW} \simeq 260$ K[5,23-29]. The charge density wave transition also manifests in our temperature-dependent tunneling spectra presented in Fig. 3**c**. The insulating gap shrinks progressively as we raise the temperature and finally a semi-metallic behavior is seen at 280 K (blue curves in Fig. 3**c**). Concurrently, tunneling spectra taken at a monolayer step edge (orange curves in Fig. 3**c**) show that the in-gap edge state, which has an enhanced d$I$/d$V$ within the insulating gap, is present throughout the range of temperatures below $T_{CDW}$. However, above $T_{CDW}$, the d$I$/d$V$ spectrum at the edge is suppressed compared to the d$I$/d$V$ spectrum away from the edge, signaling the absence of the edge state in the semimetal phase of Ta$_2$Se$_8$I. We further substantiate this observation by acquiring spectroscopic maps as the transition temperature is crossed (Figs. 3**d, e**). At 250 K, a pronounced in-gap localized state is present at the edge whereas, at 280 K, the d$I$/d$V$ at the edge is rather suppressed. These results are in accordance with a transition from the charge density wave insulator to the Weyl semimetal phase. It is worth noting that the Weyl physics in Ta$_2$Se$_8$I exists within a narrow energy window, on the order of < 20 meV, which is significantly smaller than the charge density wave gap (~500 meV). Furthermore, as demonstrated in our experiments and corroborated by our theoretical analysis below, the localization length of the edge state varies inversely with the charge density wave gap and is expected to diverge as the charge density wave gap closes in the Weyl semimetal phase (Extended Fig. 8). Consequently, it is highly improbable that the Weyl physics plays a relevant role in the existence of the edge state.

The temperature dependence reveals that the edge state is quite robust and vanishes as soon as the charge density wave gap closes above $T_{CDW}$ suggesting its connection to the charge density wave insulating gap. To gain deeper insights into the relationship between the edge state and the charge density wave, we focus on a monolayer step edge, conducting comprehensive spectroscopic mapping as summarized in Fig. 4. In Fig. 4**a,** we present the topography and corresponding representative differential conductance maps around a monolayer step edge, acquired at various bias voltages, unveiling the presence of an in-gap edge state. d$I$/d$V$ line profiles along the edge extracted from these spectroscopic maps reveal a periodic modulation with a periodicity of $\lambda_{avg} = 14.4 \pm 0.7$ nm (Figs. 4**b, d**), equivalent to $(1.43 \pm 0.07) \lambda_{CDW}$ (where $\lambda_{CDW}$ is obtained from the topography), closely matching the charge density wave period projected along the edge ($\sqrt{2} \lambda_{CDW}$). This correspondence provides robust evidence of a connection between the edge state and the charge order. Notably, the periodicity of the in-gap, edge d$I$/d$V$ oscillations, and consequently the associated wavevector obtained through a Fourier transform, remains unaltered with energy (Fig. 4**d**). Intriguingly, however, the *phase* of the charge order, $\phi$, gradually shifts with energy, which is readily apparent through the discernible shifts in the positions of the d$I$/d$V$ peaks as the bias is varied within $\Delta_{CDW}$. Note that this smooth energy-phase relation is reminiscent of the topological spectral flow of the edge modes. Figure 4**c** displays the energy-phase relation at the edge, demonstrating that the phase of the charge order at the gapless edge state vary gradually with energy solely for the bias voltages falling within the bulk energy gap ($\Delta_{CDW}$), resulting in a cumulative phase shift of $\pi$ between the bulk valence and conduction band edges. Beyond this energy range, the phase remains constant within the bulk conduction or valence bands. When investigating the d$I$/d$V$ line profiles along the same direction as in Fig. 4**b**, but away from the edge, we encounter d$I$/d$V$ oscillations sharing the identical $(1.43 \pm 0.07) \lambda_{CDW}$ periodicity (Fig. 4**e**). Two critical aspects stand out in the data of Fig. 4**e**. Firstly, there are no data points within $\Delta_{CDW}$ due to the absence of a discernible bulk d$I$/d$V$ signal at bias voltages falling within $\Delta_{CDW}$, characteristic of an insulating bulk gap. Secondly, the phase remains constant within the conduction (or valence) band as a function of energy, yet there is a $\pi$ phase difference between d$I$/d$V$ oscillations at the conduction and valence bands, creating a sharp $\pi$ phase gap in conjunction with $\Delta_{CDW}$ (illustrated in Fig. 4**f**). Notably, the $\pi$ phase difference in the charge order between the bulk conduction and valence bands can also be inferred from the data in Figs. 1**h**, **k**. Lastly, Fig. 4**g** portrays the energy-phase relation of both the edge and bulk, juxtaposed to illuminate a crucial finding from our experiments. Specifically, it illustrates that the edge state - being gapless in both energy and phase - connects the bulk conduction and valence bands that are gapped by $\Delta_{CDW}$ ($\pi$) in energy (phase). This parallels the manner in which gapless topological boundary modes connect the gapped bulk bands in terms of energy and momentum magnitude. Such a linkage between the edge state and the underlying bulk charge order gap represents an unprecedented occurrence within the realm of wide-ranging charge density wave systems — consider, for instance, the case of TaTe$_4$, where an edge state was identified[56], though no definitive association between the edge state and the charge order was established.

It is tempting to interpret the underlying quantum state of the charge density wave insulator phase in Ta$_2$Se$_8$I as a weak topological insulator composed of a stacking of two-dimensional quantum spin Hall states, with a stacking direction (equal to the vector of weak indices) perpendicular to the measured surface. However, theoretical arguments suggest that the vector of weak indices should be parallel to the ordering vector of the charge density wave[57], which is incompatible with our experimental observation. Another possibility is to have a gapped surface (and an edge state therein) due to surface reconstruction which can dimerize the pair of surface Dirac cones by breaking translation. However, the observed Bragg peaks (Fig. 2**g** and Extended Fig. 2**b**) are consistent with the ones that have been obtained with other bulk-sensitive techniques[23-29], showing no sign of additional surface reconstruction, thus contradicting this scenario. Therefore, while pointing to a topological nature of the observed edge state linked to the charge density wave phase, our experiments invite future theoretical works to understand whether it is related to the Weyl semimetal nature of the high-temperature phase.

To better understand the properties of the observed edge states and elucidate the essential physics, we construct a tight-binding model based on the nearly square iodine lattice and consider the charge density wave as a periodic potential in a sinusoidal form. Furthermore, we consider the system in a square geometry with the edges in the diagonal directions ($\sim 45^o$ with respect to the square iodine lattice). In the absence of charge density wave, the system is gapless. The charge density wave opens sizable bulk gaps at low energies of the system with an oscillation pattern, which is consistent with the bulk d$I$/d$V$ spectrum (Extended Figs. 9**c, d**). Strikingly, within the bulk gap, we find edge states which are exponentially localized at the edges. Moreover, by choosing appropriate parameters, we can obtain a bulk gap of size $\sim$ 550 meV and, accordingly, a localization length $r_0 \sim$ 1.3 nm of the edge states (Extended Figs. 9**e, f**). The localization length becomes longer as we decrease the bulk gap. The edge states exhibit a pronounced oscillation with a period $\sqrt{2}\ \lambda_{CDW}$ along the edge. Furthermore, the peak positions of the edge states along the edge shift for different biases, indicating an energy dispersion as a function of position[58]. These results are also in good agreement with our experimental observations. To understand the origin of the edge states, we analyze the decoupled wire limit where the hopping in one direction vanishes (Extended Fig. 10). In this limit, all wires are the same with identical charge density wave phases and strengths. They exhibit topological bulk gaps, which are characterized by nonzero topological numbers, and consequently in-gap edge states under open boundary conditions. As we reduce the hopping strength in one direction to zero, the bulk gaps induced by the charge density wave remain open. This indicates that the topological insulating phase of our square lattice model is continuously connected to that of the decoupled wire limit. Thus, the edge states share the same topological origin and are characterized by the same topological numbers. We provide the details and more discussions of the model calculations in Methods. While more detailed, three-dimensional *ab initio* calculations are required to fully capture the topological properties of the charge density wave phase in Ta$_2$Se$_8$I, our two-dimensional effective model, built upon the surface square lattice of I atoms, suggests a feasible and novel mechanism for the emergence of the edge state. It is essential to note, however, that conducting these *ab initio* calculations is currently highly computationally demanding due to the large charge density wave wavelength in this material.

The impact of our results is threefold. First, we visualize an in-gap edge state in $Ta_2Se_8I$ within the charge density wave gap. Going beyond the existing theories and the axion insulator interpretation of the charge density wave phase in $Ta_2Se_8I$, it indicates that the charge density wave state possibly contains a different, unique topology. Second, we find that the edge state is robust and persists up to $T_{CDW}$ ($\simeq$260 K) before disappearing quickly above $T_{CDW}$. Additionally, we show a correlation between the decay length of the edge state and the strength of the charge order, as reflected in the increase of the decay length with the diminishing charge order gap. Remarkably, while the edge d$I$/d$V$ spectra exhibit oscillations with a periodicity closely matching the charge density wave period projected along the edge, the phase of these oscillations undergoes a π shift from the conduction band edge to the valence band edge. This phase shift of the gapless edge state connects the gapped bulk conduction and valence bands in both energy and *phase*, akin to a topological bulk-boundary connectivity in energy and momentum *magnitude*, albeit distinct in its phenomenology. Although tunneling spectroscopy experiments alone cannot provide conclusive evidence of the topological nature of the edge states, these exotic behaviors collectively suggest the presence of a topologically non-trivial state, as supported by our theoretical analysis. Lastly, the charge density wave gap size surpasses the predictions from weakly-coupled mean-field theory, suggesting a potential strong coupling nature, thus opening a new avenue for investigating the interplay between charge order, strong coupling, and topology.

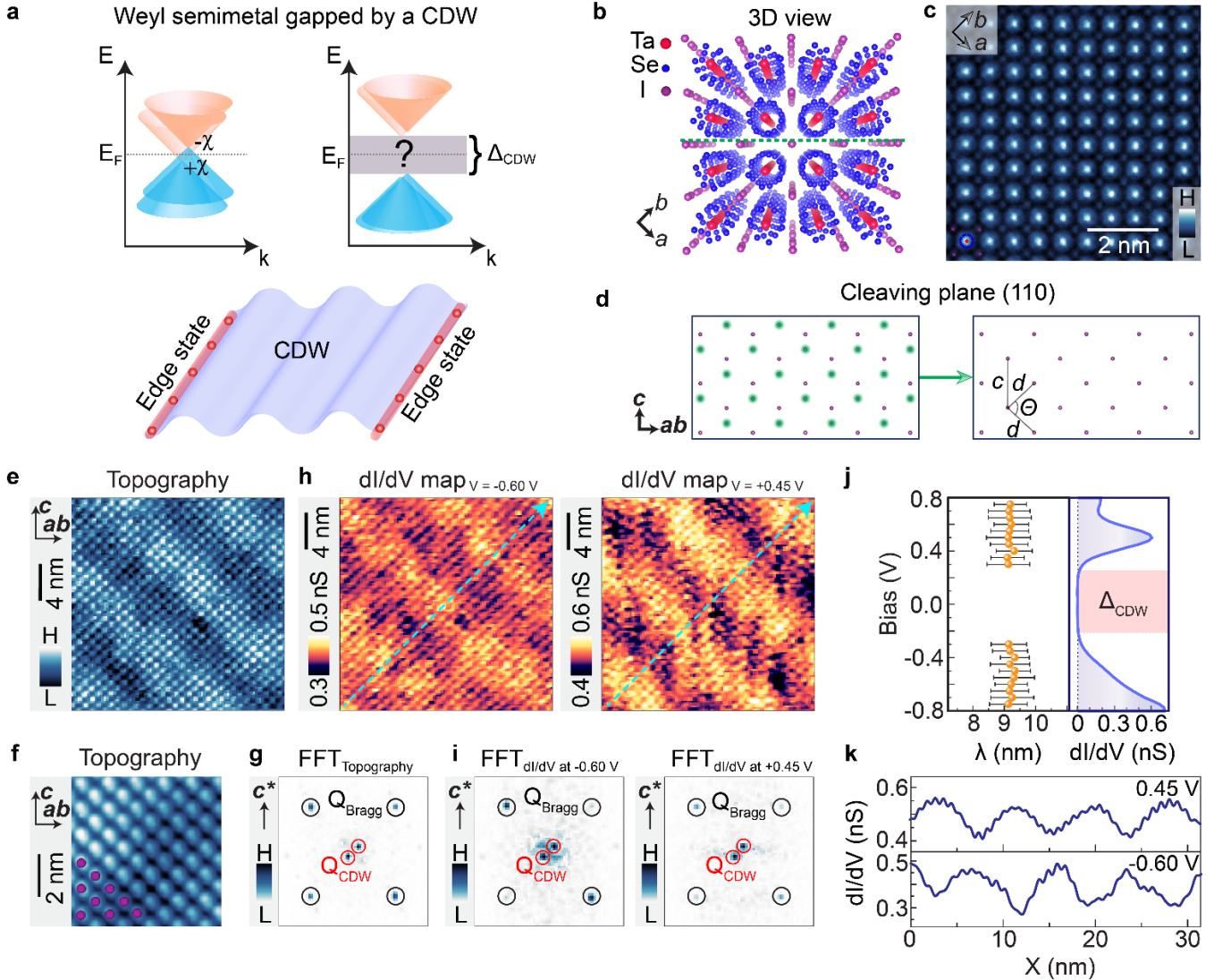

**Fig. 1: Electronic nature of charge density wave in Ta$_2$Se$_8$I. a**, Top: Schematic of phase transition from a Weyl semimetal into a charge density wave insulator[13]. Refs. [4,5,14] suggested that a Weyl semimetal can be gapped out into an axion insulator for the special case where Weyl nodes are located at the same energy. However, this does not apply to chiral crystals, where the paired Weyl fermions are at different energies[59]. Thus, the topological nature of the charge density wave in more general cases, e.g., in Ta$_2$Se$_8$I, still eludes theoretical understanding. Bottom: Real space representation of an edge state residing within the insulating charge density wave gap. **b**, 3D view of the crystal structure of Ta$_2$Se$_8$I, the green dashed line marks the cleaving plane. **c**, Scanning transmission electron microscopy image perpendicular to the growth direction, i.e., *ab* plane, showing quasi-one-dimensional nature of the crystal. **d,** Left: Schematic of sample cleaving plane (110) containing I atoms, 50% of whose are removed (denoted with green shadows). Note, $\overrightarrow{ab} \parallel (\vec{a} + \vec{b})$. Right: the resulting sample plane after the 50% I atoms are removed. Interatomic distances along different directions ($d \simeq 9.5$ Å, $c \simeq 13$ Å, $\Theta \simeq 86\pm0.5°$) are marked. **e**, Atomically resolved topographic image of the (110) plane exhibiting a clear charge density wave with $\lambda_{CDW} = 9.1 \pm 0.6$ nm ($V_{gap} = -1.0$ V, $I_t = 50$ pA). **f**, Enlarged view of the topographic image showing a clear lattice of iodine

atoms. **g**, Fast Fourier Transform of the topographic image in panel **e**, displaying well-developed charge density wave peaks along with the Bragg peaks. **h**, d$I$/d$V$ maps using $V_{gap}$ = -0.60 V (left) and $V_{gap}$ = 0.45 V (right), $I_t$ = 0.3 nA, $V_{mod}$ = 10 mV, demonstrating a reversal of the spectroscopic contrast of the charge density wave as the tip-sample bias is tuned from the valence to the conduction band. **i**, Fourier transforms of the d$I$/d$V$ maps in panel **h** acquired at $V_{gap}$ = -0.60 V (left) and $V_{gap}$ = 0.45 V (right), showcasing the charge density wave peaks. **j**, Left: Energy dependence of the wavelength of the charge density wave derived from d$I$/d$V$ maps. Right: The averaged differential spectrum, replicated from Extended Fig. 4b, showcasing a large insulating gap (the dotted vertical line indicates zero differential conductance). **k**, d$I$/d$V$ line profiles derived from linecuts along the $\vec{q}_{CDW}$ direction using the data presented in the panel **h**, showcasing a distinct spectroscopic contrast inversion of the charge density wave as the bias is changed from -0.60 V to 0.45 V. The line profiles are acquired from the same spatial location, marked with dashed cyan lines in panel **h**; the direction is marked with arrows. Tunneling junction set-up: $V_{set}$ = -0.8 V, $I_{set}$ = 0.4 nA, $V_{mod}$ = 10 mV. The data in panels **e-k** are taken at $T$ = 160 K.

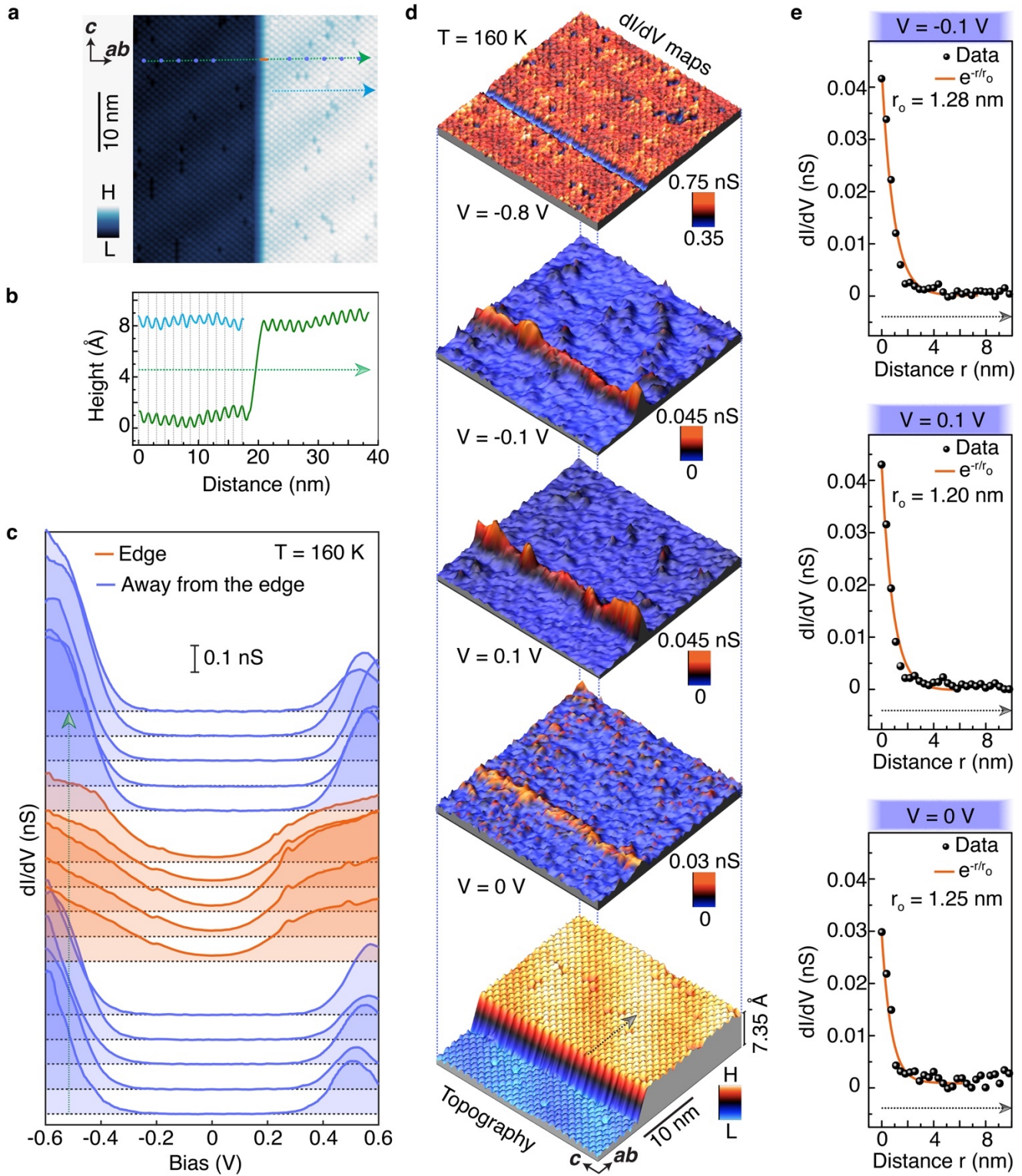

**Fig. 2: Observation of an edge state within the charge density wave gap. a**, Atomically resolved topography of an atomically sharp monolayer step edge ($V_{gap}$ = -1.0 V, $I_t$ = 50 pA). **b**, The corresponding height profile, taken perpendicular to the *c*-axis direction, exhibits uniform atomic correlations perpendicular to the step edge over a large distance. The vertical dotted lines denote the atomic positions perpendicular to the $c$-axis. **c**, Tunneling spectroscopy taken at $T = 160$ K, revealing an insulating gap away from the edge and a pronounced in-gap state on the edge. Orange and blue curves represent the differential spectra taken at different positions (marked on the panel **a**) at the edge and away from the edge, respectively. Spectra are offset for clarity. Dotted horizontal lines denote zero differential conductance. Tunneling junction set-up: $V_{set} = -0.8$ V, $I_{set} = 0.4$ nA, $V_{mod} = 10$ mV. **d**, d$I$/d$V$ maps taken at 160 K at different bias voltages (corresponding topography is shown in the bottom panel). d$I$/d$V$ maps taken within the insulating gap ($V = 0$ mV and $\pm 100$ mV) reveal a pronounced in-gap edge state. Tunneling junction set-up for d$I$/d$V$ maps: $V_{set} = -0.8$ V, $I_{set} = 0.3$ nA, $V_{mod} = 10$ mV. **e**, Intensity distribution of differential conductance taken at 0 mV and $\pm 100$ mV away from the step edge on the crystal side. The corresponding location is marked on the topographic image in panel **d** with a black dashed line; the direction is marked with an arrow. The orange curves show the exponential fitting of the decay of the state away from the edge.

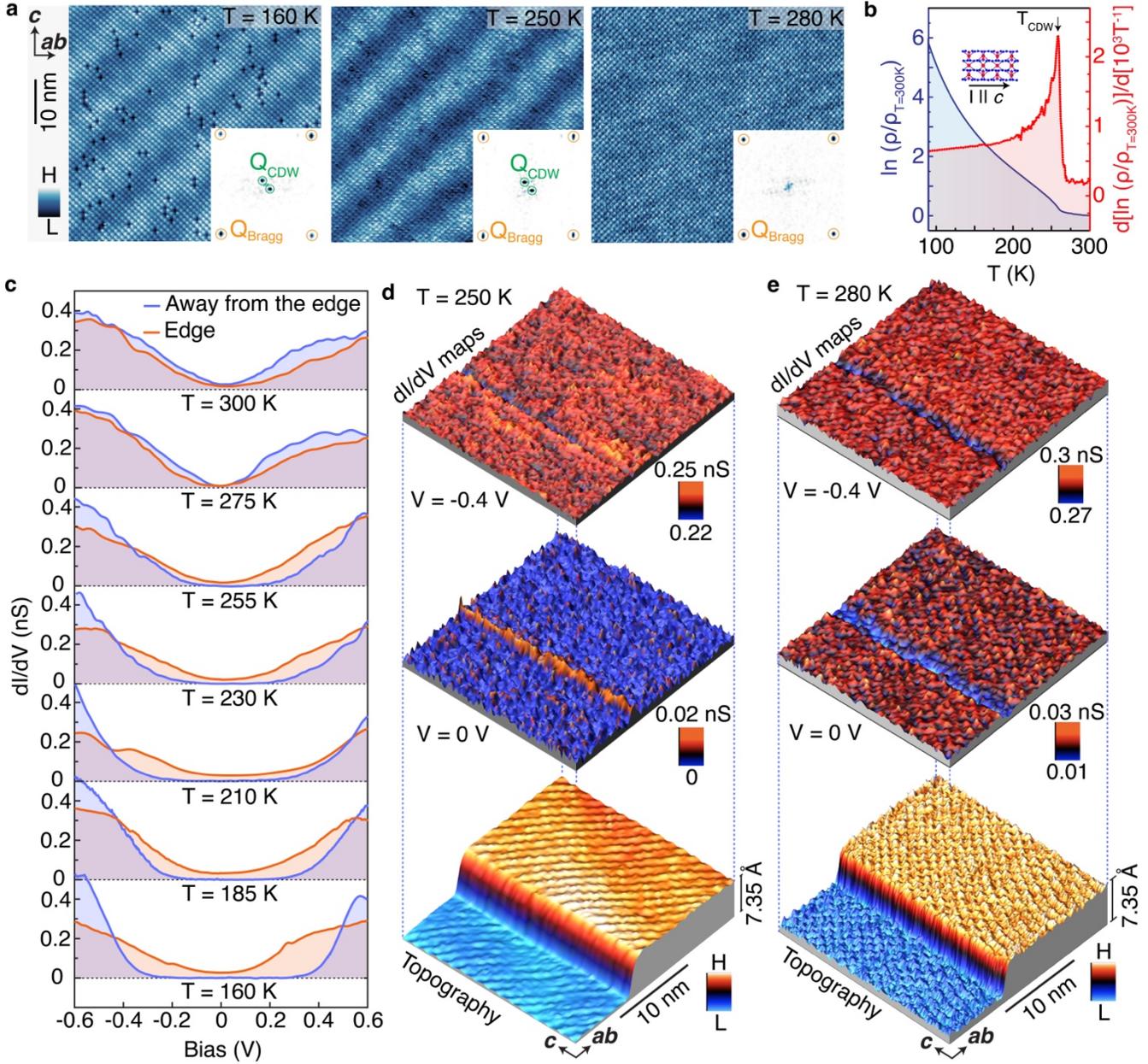

**Fig. 3: Temperature dependence of the charge density wave and its in-gap edge state. a,** Topography images and the corresponding Fourier transforms (shown in the inset) taken at three different temperatures highlighting the temperature evolution of the charge density wave ($V_{gap}$ = -1.0 V, $I_t$ = 50 pA). At $T$ = 280 K, the charge density wave and its Fourier transform peaks are not detected. **b,** Resistivity vs temperature data showing a kink near 260 K signaling the charge density wave transition which corroborates our observation. The red curve represents the plot of the derivative of the natural logarithm of the ratio of $\rho$ to $\rho_{T=300\ K}$ with respect to $\frac{10^3}{T}$. **c,** Temperature-dependent differential spectra taken at the edge and away from the edge, denoted with orange and blue curves, respectively. Spectra are offset for clarity for different temperatures. In agreement with the topography, the charge density wave gap and the associated in-gap edge state are not visible above the transition temperature. Tunneling junction set-up: $V_{set}$ = -0.8 V, $I_{set}$ = 0.4 nA, $V_{mod}$ = 10 mV. **d** and **e,** Topography and corresponding differential conductance maps

taken on a monolayer step edge at $T = 250$ K (panel **d**) and 280 K (panel **e**) capturing the temperature dependence of the edge state. At $T = 250$ K, the edge state is still visible whereas, at $T = 280$ K, the edge state is absent. Tunneling junction set-up for d$I$/d$V$ maps: $V_{set} = -0.8$ V, $I_{set} = 0.3$ nA, $V_{mod} = 10$ mV.

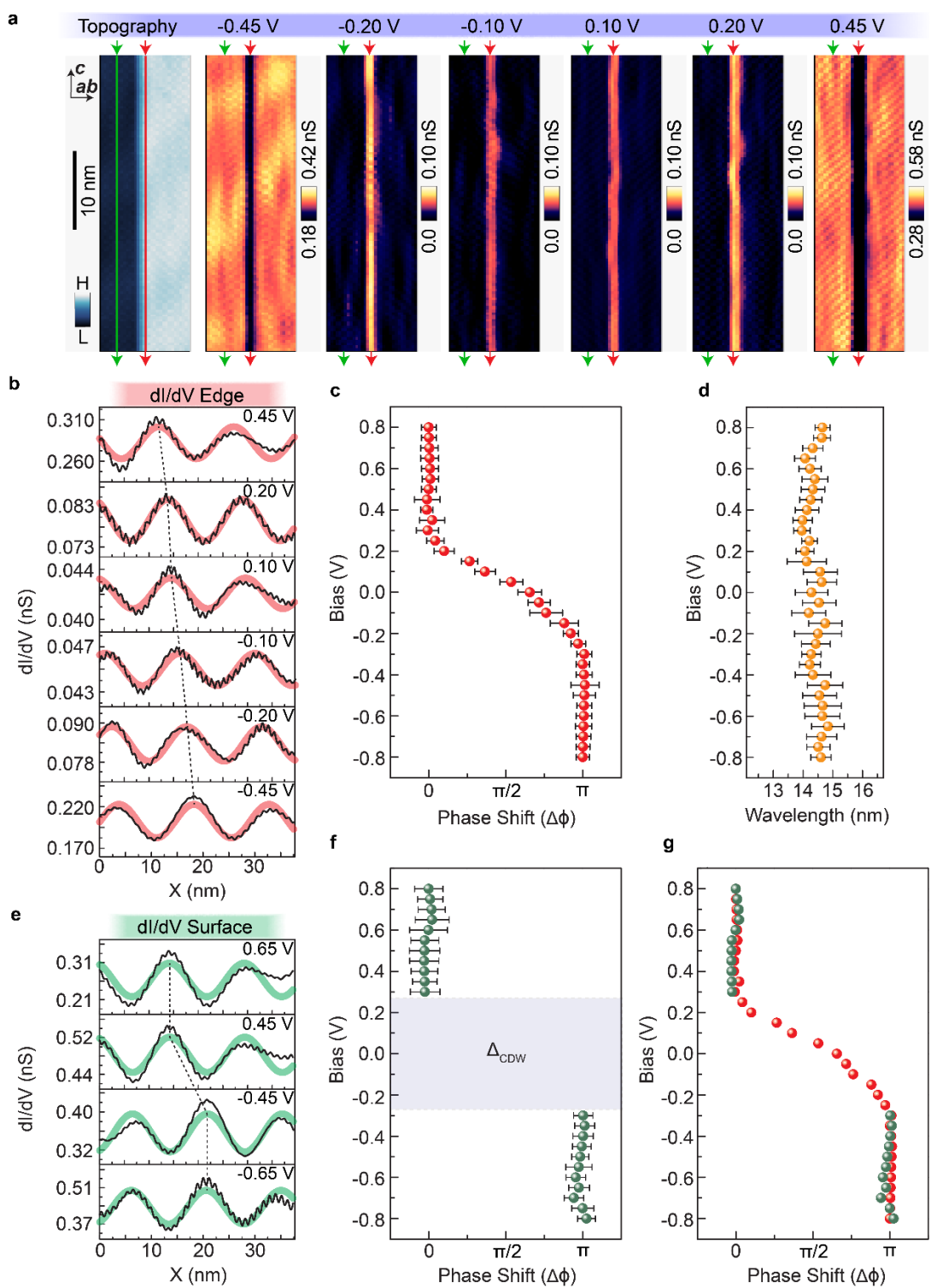

**Fig. 4: Gapless edge mode connecting the gapped bulk (and surface) conduction and valence bands in energy and phase. a,** Topography and corresponding representative differential conductance maps around a monolayer

step edge shown for six different bias voltages, revealing the presence of an in-gap edge state. The red arrows indicate the location of the edge in the *ab* axis. Tunneling junction set-up for d$I$/d$V$ maps: $V_{set}$ = -0.8 V, $I_{set}$ = 0.3 nA, $V_{mod}$ = 10 mV. All data are taken at $T$ =160 K. The charge density wave in this specific sample displays a periodicity of $\lambda_{CDW} = 10.1 \pm 0.3\ nm$, deduced from topographic imaging of a region located away from the edge. **b,** d$I$/d$V$ line profiles (averaged over an atom along the edge marked with a red vertical line in the topographic image in panel **a**) along the edge shown for six biases ranging from -0.8 V to 0.8 V, exhibiting a periodic modulation which can be fitted with $A\sin\frac{2\pi}{\lambda}(x - \phi)$; fitting parameters are $\lambda$ and $\phi$ which denote wavelength and phase shift, respectively. Averaging $\lambda$ obtained from the fitted sinusoids of all such acquired d$I$/d$V$ curves, we obtain $\lambda_{avg}$ = 14.4 $\pm$ 0.7 nm. The red curves denote fitting of $A \sin\frac{2\pi}{\lambda_{avg}}(x - \phi)$ applied to d$I$/d$V$ data at each bias. **c,** Deduced phase shift $\Delta\phi = \phi - \phi_{V=0.8V}$ of the d$I$/d$V$ oscillations (obtained from the fit marked with red curves) akin to what is shown in panel **b**, plotted as a function of bias. As also visualized through the shift of the peak locations with varying bias in panel **b**, the phase shifts with energy solely for the bias voltages falling within the bulk energy gap ($\Delta_{CDW}$), resulting in a cumulative phase shift of $\pi$ between the bulk valence and conduction band edges. Beyond this energy range, the phase remains constant within the bulk conduction or valence bands. **d,** Deduced periodicity ($\lambda$) of the edge d$I$/d$V$ oscillations plotted against the bias voltage, highlighting an energy-independent $\lambda$ and consequent wavevector. Note that $\lambda_{avg}$ aligns well with the charge density wave wavelength projected along the edge, i.e., $\sqrt{2}\ \lambda_{CDW}$. **e,** d$I$/d$V$ line profiles (averaged over an atom along the *c*-axis direction marked with a green vertical line in the topographic image in panel **a**) away from the edge shown for four biases ranging from -0.8 V to 0.8 V, exhibiting a periodic modulation which can be fitted with $A\sin\frac{2\pi}{\lambda}(x - \phi)$; fitting parameters are $\lambda$ and $\phi$ which denote wavelength and phase shift, respectively. Averaging $\lambda$ obtained from the fitted sinusoids of all such d$I$/d$V$ curves, we obtain $\lambda_{avg} = 14.2 \pm 0.6\ nm$. The green curves denote fitting of $A \sin\frac{2\pi}{\lambda_{avg}}(x - \phi)$ applied to d$I$/d$V$ data at each bias. **f,** Deduced phase shift $\Delta\phi = \phi - \phi_{V=0.8V}$ of the d$I$/d$V$ oscillations (obtained from the fit marked with green curves) akin to what is shown in panel **e**, plotted as a function of bias. There are no data points within $\Delta_{CDW}$ due to the absence of a discernible bulk d$I$/d$V$ signal at bias voltages falling within $\Delta_{CDW}$, characteristic of an insulating bulk gap. As visualized through tracking the peak locations with changing bias in panel **e**, the phase does not change within both the conduction and valence bands as a function of energy, yet there is a $\pi$ phase shift between d$I$/d$V$ oscillations acquired at the conduction and valence bands, creating a sharp $\pi$ phase gap, present in conjunction with $\Delta_{CDW}$. **g,** Energy vs. phase plots of the edge and the bulk shown in panels **c** and **f**, respectively, plotted together. The plot highlights the presence of a gapless edge state connecting the gapped (by $\Delta_{CDW}$ in energy and $\pi$ in phase) bulk conduction and valence bands in energy and phase, bearing resemblance to the topological spectral flow of the boundary modes.

## Methods:

**Ta₂Se₈I single-crystal growth and characterization**

Tantalum powder, -325 mesh, Puratronic®, 99.97% (metals basis), Iodine lump, ultra-dry, (99.999%, metals basis), Selenium powder, -200 mesh, 99.999% (metals basis), were purchased from Alfa Aesar and used for synthesis without further purification. Single crystals of Ta₂Se₈I were grown using a chemical vapor transport method with Ta, Se, and I as starting materials. Ta: Se: I = 2:8:1 composition ratio was taken and sealed in an evacuated quartz tube. The ampoule was placed into a pre-heated dual-zone tube furnace with a temperature gradient of 500 to 400 °C and the educts in the hot zone for two weeks. After the growth period, a few mm-long needle shape (with aspect ratio of approximately 10) crystals had formed in the cold zone.

We conducted structural analysis on the cleaved Ta₂Se₈I sample surface using electron diffraction measurements. In Extended Fig. 1**b**, we display the scanning electron microscopy image of the thin lamella, which was obtained through focused ion beam cutting from the cleaved surface that was required for performing the electron diffraction experiments. Extended Fig. 1**c** provides the electron diffraction pattern acquired at $T = 290$ K. Our electron diffraction analysis confirms that the cleaved surface corresponds to the (110) plane.

**Scanning tunneling microscopy**

Single crystals were cleaved mechanically in situ in ultra-high vacuum conditions ($< 5 \times 10^{-10}$ mbar), and then immediately inserted into the microscope head. Tunneling junction set-up parameters for topography and spectroscopy measurements are noted in figure captions. Topographic images in this work were taken in the constant current mode. Tunneling conductance spectra were obtained with a commercial W tip using the standard lock-in amplifier technique with a lock-in amplifier frequency of 977.77 Hz. For variable temperature measurement, we first withdrew the tip from the sample, and then raised the temperature and stabilized the temperature for 12 hours, after which we reapproached the tip to the sample to perform tunneling experiments.

**Energy gap determination**

In this section, we outline a procedure that we deployed to determine the value of the spectroscopic energy gap from d$I$/d$V$ spectra. The key idea is to identify the edges of the energy gap, where a non-zero d$I$/d$V$ signal can be distinguished from the zero d$I$/d$V$ noise signal within the gapped region of the spectrum[60,61]. The following are the series of steps to determine the energy gap from a single d$I$/d$V$ curve (we refer the reader to Extended Fig. 5).

1. Initially, we selected a bias range where the d$I$/d$V$ signal takes the form: d$I$/d$V = 0 + \xi$, where 0 represents the mean d$I$/d$V$ value within the spectroscopic gap and $\xi$ represents the noise in the d$I$/d$V$ signal. This was done for some interval $V \in [V_1, V_2]$, while avoiding the edges of the spectroscopic gap.

2. We then calculated the standard deviation of the d$I$/d$V$ noise signal within the gap, $\sigma = \sqrt{\overline{\xi^2}}$. By determining $\sigma$ from the gapped segment of the d$I$/d$V$ curve, we established a noise floor of the d$I$/d$V$ signal.

3. We assume a Gaussian distribution of $\xi$ and define $\Gamma = 2.36\sigma$, which represents the full width at half maximum. Two Gaussian signals can be distinguished if the difference between the means of the signals is greater than the full width at half maximum. Thus, we set $\Gamma$ as the instrumental resolution of the d$I$/d$V$ signal.

4. A non-zero d$I$/d$V$ signal outside of the gapped region can only be detected when d$I$/d$V > \Gamma$ with our instrumental resolution. Therefore, we set $\Gamma$ as a threshold value for identifying the edges of the energy gap.

5. By numerically solving the equation $dI/dV = \Gamma$, we determined the intersections $V_a, V_b$ of the threshold value with the $dI/dV$ spectroscopic curve. $eV_a$ and $eV_b$ represent the energies of the gap edge below and above $E_F$, respectively. Finally, we calculated $\Delta = eV_b - eV_a$ to obtain the value of the spectroscopic energy gap.

**Angle-resolved photoemission spectroscopy measurement**

The angle-resolved photoemission spectroscopy measurement was performed at SIS beamline of Swiss Light Source (SLS), Switzerland. The energy and momentum resolution were below 20 meV and 0.02 Å$^{-1}$, respectively. The Fermi level was determined by measuring a polycrystalline gold which was electrically grounded with the measured sample.

**Charge density wave wavevector calculation**

Here we discuss how we obtain the charge density wave wavevector from the scanning tunneling microscopy data. As stated in the main text and detailed in our extended scanning tunneling microscopy data, the charge density wave retains its phase in the (110) plane even if it encounters a step edge. This observation suggests that the charge density wave wavevector ($\vec{q}_{CDW}^{3D}$) lies in the (110) plane. Therefore, $\vec{q}_{CDW}^{3D}$ can be expressed as: $\vec{q}_{CDW}^{3D} = (\eta \frac{2\pi}{a}, -\eta \frac{2\pi}{a}, \pm\delta \frac{2\pi}{c})$ or $\vec{q}_{CDW}^{3D} = (-\eta \frac{2\pi}{a}, \eta \frac{2\pi}{a}, \pm\delta \frac{2\pi}{c})$. Consequently, $\vec{q}_{CDW}^{3D}$ projected on (110) plane has the following components: $\vec{q}_{CDW}^{(110)} = (q_{ab}, q_c) = (\pm\sqrt{2}\eta \frac{2\pi}{a}, \pm\delta \frac{2\pi}{c})$. We can obtain $q_{ab}$ and $q_c$, i.e., $\vec{q}_{CDW}^{(110)}$ directly from the Fourier transform of the topographic image (Extended Fig. 2). In fact, $(q_{ab}, q_c)$ is precisely the location of the charge density wave peak in the Fourier transform image (Extended Fig. 2**b**). Thus, we determine the $\eta$ and $\delta$ components of $\vec{q}_{CDW}^{3D}$, leading to $\vec{q}_{CDW}^{3D} \sim (0.045 \frac{2\pi}{a}, -0.045 \frac{2\pi}{a}, 0.083 \frac{2\pi}{c})$.

We would also like to add that the directions of *ab* and *c* axes can be directly determined from the Fourier transform of the topographic image (Extended Fig. 2**b**). Because the lattice constants along *ab* ($\simeq$13.8 Å) and *c* ($\simeq$13.0 Å) axes are slightly different, the *q*-vectors of the 2$^{nd}$ order Bragg peaks have different magnitudes- $|\vec{q}_{ab}^{2^{nd}Bragg}| < |\vec{q}_c^{2^{nd}Bragg}|$. Therefore, a relative comparison of these $\vec{q}$ vectors in the Fourier transform allows us to discern *ab* and *c* axes in the topographic image.

**Phase continuity of the charge density wave across step edge**

Here we delineate how the phase of the charge density wave evolves as it encounters an atomic step edge. We start by showing in Extended Fig. 3**a** a detailed analysis of the charge density wave phase on the two sides of the monolayer step edge. We take two line profiles, one on each side of the monolayer step edge, along the charge density wave direction, as elucidated in Extended Fig. 3**a**. We then fit a sinusoidal curve to the profiles to estimate the phase of the charge density wave on the two sides (Extended Fig. 3**b**). We find no apparent phase shift between the two fitted sinusoidal curves. This observation is consistent with the charge density wave wavevector lying completely in (110) plane.

It is worth noting that all step edges we have observed in numerous (more than 30) Ta$_2$Se$_8$I samples were oriented along the *c*-axis. Because of the crystal structure of Ta$_2$Se$_8$I, it is extremely difficult to find edges other than those along the *c*-axis direction (Extended Fig. 1).

**Energy gap map showing the charge density wave modulations**

As discussed in the main text, the energy gap usually correlates with the electronic nature of the sample only, irrespective of the tip. Extended Figs. **4d, e** delineate our energy gap map measurement. The energy gap map (Extended Fig. **4e**) exhibits charge density wave modulations akin to the corresponding topographic image (Extended Fig. **4d**) [29,37]. Fourier transforms of the topography and the gap map reveal similar peaks corresponding to the charge density wave. The presence of such charge density wave peaks in the energy gap map is consistent with the electronic nature of the charge density wave.

**Absence of surface state probed via angle-resolved photoemission spectroscopy**
In Extended Fig. 6, we show angle-resolved photoemission spectroscopy data taken at the charge density wave phase of $Ta_2Se_8I$. All the observed states (shown in panel **a**) match prior first-principles calculation of the bulk states[5,20] and experimental results[20]. Notably, there is no spectral weight inside the bandgap (labeled in panel **b**). This observation indicates the absence of a surface state.

**Comparison with the prior scanning tunneling microscopy report**
Here we compare our data with a recent scanning tunneling microscopy study of $Ta_2Se_8I$ by Huang et al.[21]. Huang et al.[21] focused on charge density wave edge dislocations found on the uniform surface with no height discontinuities. They examined the tunneling spectroscopies at and very close to the charge density wave edge dislocations (see Figs. 3**c, d** of ref. [21]). No in-gap states were observed at such locations[21]. In stark contrast, we primarily examine the pristine, atomically sharp, monolayer step edge. Our scanning tunneling microscopy measurements reveal a clear in-gap state at the atomically sharp step edge. We do a systematic study of the edge state as a function of temperature showing a clear correlation with the charge density wave gap. Furthermore, we experimentally uncovered a phase dispersion in the edge state, illustrating how this gapless edge state connects the gapped bulk conduction and valence bands in terms of energy and phase. These findings highlight a distinctive topological connection between the edge state and the charge density wave energy gap.

Continuing our comparison with ref. [21] data, the charge density wave period in our data is $<\lambda_{CDW}> = 10.4 \pm 1.6$ nm, which is smaller than $\lambda_{CDW} \sim 17\pm1$ nm reported by Huang et al.[21]. We note that our charge density wave period matches more closely with $\lambda_{CDW} \sim 11$ nm obtained in prior X-ray and neutron scattering experiments[23-29]. Even though the differential spectra in our work and ref. [21] share similar features– a shoulder at negative bias and a broad peak at positive bias, there is a quantitative difference in the energy gap value. Huang et al.[21] found a gap size of ~200 meV; this is smaller than our measured value of ~500 meV. We add that the gap size we obtain agrees with the prior photoemission spectroscopy measurements[14,20,33], but is larger than what was observed in transport experiements[5,20]. While there is no qualitative inconsistency between the two scanning tunneling microscopy works, the quantitative discrepancies in $\lambda_{CDW}$ and the gap size may stem from different crystal synthesis conditions and the difference in the amount of the I vacancies present in the sample.

In ref. [21] no spectroscopic evidence of charge modulation in d$I$/d$V$ maps or spatial profiles is presented. These limitations could be attributed to the larger presence of I-vacancies in the samples used in ref. [21], and restricts the ability to draw definitive conclusions about the nature of the charge density wave. In stark contrast, our work provides compelling evidence for the electronic nature of the charge density wave in $Ta_2Se_8I$. We present several key observations that strongly support this conclusion:

1. We perform spectroscopic mapping of the charge density wave in $Ta_2Se_8I$ and present corresponding d$I$/d$V$ maps at -0.60 V and +0.45 V in Fig. 1**h**. The Fourier transform of these maps clearly exhibits peaks corresponding to the charge density wave wavevector, which align with the peaks observed in the Fourier transform of the topography. Notably, we observe a distinct spectroscopic contrast switch in the charge order as the bias is varied from -0.60 V to +0.45 V. This behavior is typically associated with the electronic nature of the charge order[30-32].

2. We acquire high-quality spectroscopic data in a pristine region of $Ta_2Se_8I$ containing minimal I-vacancies and analyze the spatial dependence of the charge density wave insulating gap (Extended Fig. 4). The spatially resolved tunneling spectra along the charge density wave wavevector direction reveal a pronounced modulation of the insulating gap with a periodicity that closely matches the topographic periodicity (Extended Fig. 4c). Additionally, in Extended Figs. 4d, e, we present an energy gap map, demonstrating the modulation of the charge density wave gap. The position of the charge density wave wavevector peaks in the Fourier transform of the gap map corroborates the peaks observed in the Fourier transform of the topography. Insofar as an energy gap map serves as a direct probe of the electronic nature of the sample, our observation of charge density wave gap modulation provides compelling evidence for the electronic nature of the charge density wave in $Ta_2Se_8I$.

Therefore, our spectroscopic measurements of the charge density wave in $Ta_2Se_8I$ provide compelling evidence for its electronic nature.

**Theoretical discussion on the underlying topology of the boundary modes in a charge density wave insulator**

*Description of the Model and topological edge states*: As stated in the main text, the cleaved $Ta_2Se_8I$ surface, i.e., the (110) plane consists of iodine atoms, forming a nearly square lattice with step edges in the diagonal direction ($\sim 45°$ with respect to the nearly square iodine lattice). The charge density wave has a single vector which is parallel to one direction of the iodine lattice, say the $y$-direction (as sketched by Extended Fig. 9a). To mimic these features, we construct a two-orbital minimum model with a charge density wave modulation on a square lattice described by $H = H_0 + H_{CDW}$. Here, $H_0$ is the bare electron Hamiltonian given by

$$H_0 = \sum_{\mathbf{r}} (t_x \Psi^\dagger_{\mathbf{r}+\hat{\mathbf{x}}} \sigma_z \Psi_\mathbf{r} + t_y \Psi^\dagger_{\mathbf{r}+\hat{\mathbf{y}}} \sigma_z \Psi_\mathbf{r} + h.c.),$$

where $\Psi^\dagger = (c^\dagger_{1,\mathbf{r}}, c^\dagger_{2,\mathbf{r}})$ with $c^\dagger_{o,\mathbf{r}}$ creating an electron with orbital $o \in \{1,2\}$ at position $\mathbf{r} = (x, y)$. $\hat{\mathbf{x}}$ and $\hat{\mathbf{y}}$ are the nearest-neighbor vectors in the $x$- and $y$-directions (denoted as $\mathbf{d}_1$ and $\mathbf{d}_2$ in the figures), respectively; $t_x = t_y = t$ is the hopping strength between nearest neighbor sites; $h.c.$ represents the Hermitian conjugate counterpart; and $(\sigma_x, \sigma_y, \sigma_z)$ are the Pauli matrices for the orbital degrees of freedom. Note that in the absence of charge density wave, the model is gapless.

The charge density wave modulation can be modeled as a spatially periodic local potential

$$H_{CDW} = V \sum_{\mathbf{r}} \cos(Qy + \phi) \Psi^\dagger_\mathbf{r} M \Psi_\mathbf{r},$$

where $V$ is the charge density wave strength, $Q$ is the charge density wave vector which is related to the wavelength $\lambda'$ as $Q = 2\pi/\lambda'$, and $\phi$ is the charge density wave phase. We consider a charge density wave modulation characterized by a matrix $M \in \{\sigma_x, \sigma_y\}$ which opens bulk gaps at low energies, as we will see below. Without loss of generality, we consider $M = \sigma_x$. We note that the charge density wave on the surface is likely induced through proximity to the strongly interacting Ta-Se chains inside the bulk. Thus, its strength $V$ may be comparable to or even stronger than the weak coupling $t$ between the surface iodine atoms. We ignore spin-orbit coupling since its energy scale is much smaller compared to that of the charge density wave in the system. Adding additional onsite energy modification (characterized by a matrix $M \in \{\sigma_0, \sigma_z\}$) does not change our main results qualitatively.

The atomic step edges of the cleaved Ta$_2$Se$_8$I samples are along the diagonal direction with respect to the iodine lattice. Therefore, we model the system in a square geometry with 120 lattice sizes in both $x$- and $y$-directions and with the edges in the diagonal directions, as shown in Extended Fig. 9a. We choose the lattice constant $d = 0.95$ nm and other parameters $\lambda' = 18d$, $t_x = t_y = t = 1.5$ eV, $V$=1.2 eV and temperature $T = 200$ K for simulations. Extended Fig. 9 presents the numerical results. We find that the charge density wave potential induces two bulk gaps in the system, which are of size $\Delta_{\text{CDW}} \simeq 0.48$ eV and around energies $\varepsilon_\pm \simeq \pm 0.7$ eV, respectively (Extended Fig. 9b). Note that the energy regime of the 2D lattice model is $(-4t, 4t)$, much larger than $\Delta_{\text{CDW}}$ and $\varepsilon_\pm$. We focus on one gap at $\varepsilon_-$ and set the Fermi energy to be $E_F = \varepsilon_-$ in the following. As shown in Extended Fig. 9c, we calculate the local density of states (LDOS) as a function of position $y$ inside the bulk (along the red dashed line in Extended Fig. 9a). Clearly, the LDOS shows a bulk gap around $E_F$ that varies periodically with $y$. Note that the LDOS depends only on the absolute value of the charge density wave potential. Thus, we obtain a period in the LDOS $\lambda = \lambda'/2 \simeq 9d$, which is close to the $\lambda_{\text{CDW}} \simeq 9.1 \pm 0.6$ nm observed in the experiment. Moreover, we find that the LDOS maxima (minima) at the valence band correspond to the LDOS minima (maxima) at the conduction band, as shown in Extended Figs. 9d-i, d-ii. These results are consistent with the experimental data presented in Fig. 1.

Strikingly, for any phase $\phi$, we observe in-gap edge states within the bulk gap. Their wavefunctions are exponentially localized at the edge with a small localization length $r_0 \simeq 1.6d = 1.52$ nm, as shown in Extended Fig. 9e, which are in contrast with the bulk continuum states where the wavefunctions are distributed throughout the system (Extended Figs. 9d-i, d-ii). If we decrease the bulk gap, we find longer localization lengths of the edge states. Furthermore, the edge wavefunctions exhibit a pronounced oscillation with period $\sqrt{2}\lambda$ along the edge direction. As we vary the energy inside the bulk gap, the peaks of the edge wavefunctions shift along the edge, indicating an energy dispersion as a function of position along the edge[58]. These features are also in good agreement with our experimental observations. We stress that these main features, such as the charge density wave-induced bulk gap, the exponentially localized in-gap edge states, the periodic oscillations and energy dispersion as a function of position along the edge, are quite general. They appear for any values of $\phi$, $V/t_{x,y}$, and $\delta$, and any wavelengths $\lambda' \geq 3a$. Finally, we note that while the energies of the edge states are discrete within the bulk gap, the energy gaps of edge states can be significantly reduced by additional perturbations, such as next-nearest-neighbor hopping, net average charge density wave potential $V_0 \ll V$ (i.e., $V(\mathbf{r})$ varies from $-V + V_0$ to $V + V_0$ when moving in space), enhancement or reduction of bonding (hopping) between atoms close to the boundary, and local potential at the boundary (Extended Fig. 11). Therefore, at high temperatures, the edge states can be observed throughout the bulk gap, thus consistent with our experiments.

***Topological invariant:*** The in-gap edge states are of topological origin. To understand this, it is constructive to consider the $t_x = 0$ limit in which the system can be thought of as decoupled parallel wires that extend in the $y$-direction. All wires are identical with the same charge density wave strength and phase. Thus, we can analyze each 1D wire separately. We consider periodic boundary conditions in the $y$-direction. Due to the super-periodic charge density wave potential with a long wavelength $\lambda'$, the spectrum of the wire is split into multiple sub-bands in the reduced Brillouin zone. We find that the charge density wave generates two bulk gaps of size $V$ around energies $\varepsilon_\pm \approx \pm t_y/\sin(\pi a/\lambda')$, respectively. Here, we have assumed a commensurate wavelength with rational $\lambda'/a$ for simplicity. Note that the system is periodic in $\phi$ and momentum $k_y$. A topological characterization of the system can be obtained in terms of Chern numbers defined in the $k_y$ and $\phi$ space, as shown in ref. [58]. Specifically, for each spectral gap, the Chern number can be obtained as

$$\nu = \int_{-\pi/\lambda_y}^{\pi/\lambda_y} dk_y \int_{-\pi}^{\pi} d\phi \, Tr\left[\partial_{k_y} A_\phi - \partial_\phi A_{k_y}\right],$$

where $A_j$ with $j \in \{k_y, \phi\}$ is the non-Abelian Berry connection $A_j = i\Phi^\dagger \partial_j \Phi$ which is a square matrix and based on the multiplet of eigenstates with energy below the gap in question $\Phi = (|\psi_1\rangle, |\psi_2\rangle, \ldots |\psi_m\rangle)$[58]. Explicitly, we find the Chern numbers $\nu = \pm 2$ for the two gaps, respectively. Note that these Chern numbers are independent of $V/t_y$ and rational $\lambda'/a \geq 3$. They are defined in the combined $k_y$ and $\phi$ space, different from the conventional Chern numbers defined in a full momentum space for Chern insulators, and thus do not require breaking of time-reversal symmetry. The nonzero Chern numbers indicate the appearance of in-gap edge modes when open boundaries are imposed to the wires in the $y$-direction. This is clearly confirmed by our numerical calculations in Extended Fig. 10. The edge states have their energies within the bulk gaps, and their wavefunctions are exponentially localized at the edges.

Finally, it is important to note that by increasing the coupling $t_x$ from 0 to $t$, the bulk gaps of the system remain open. This indicates that the charge density wave insulator on the square lattice with $t_x = t_y$ that we considered for the (110) plane of Ta$_2$Se$_8$I is continuously connected to that in the decoupled wire limit with $t_x = 0$. Therefore, the edge states of the model share the same topological origin as those of the decoupled wire model and is characterized by the same Chern numbers. It is worth noting that the Chern number counts the number of edge states (equivalently, the number of LDOS peaks along the edge) in a wavelength $\lambda'$ of the charge density wave potential. Since the period $\lambda$ of LDOS is half the wavelength of the charge density wave potential, i.e., $\lambda = \lambda'/2$, we expect that each period of LDOS exhibits a peak value. This result is also consistent with our experiments.

**Localization length of the charge density wave edge states**
The localization length of edge states is independent of the charge density wave wavelength $\lambda$ but determined by the charge density wave gap $\Delta_{CDW}$. To elucidate this property, we consider a simple one-dimensional two-orbital model

$$H = t \cos(k_x c) \sigma_z$$

with the charge density wave potential

$$H_{CDW} = V \cos\left(\frac{2\pi}{\lambda'} x + \phi\right) \sigma_s,$$

where $c$ is the lattice constant, $V$ is the potential strength, and $\phi$ is the phase shift. $(\sigma_x, \sigma_y, \sigma_z)$ are the Pauli matrices for the orbital degrees of freedom. Without charge density wave, the model is gapless. The charge density wave potential $H_{CDW}$ with $\sigma_s \in \{\sigma_x, \sigma_y\}$ opens two bulk gaps at low energies (i.e., close to the band center $E = 0$). For concreteness and without loss of generality, we consider $\sigma_s = \sigma_x$. Moreover, to simulate Ta$_2$Se$_8$I, we take the parameters $c = 1.3$ nm, $\lambda = 10c$, $t = 1.2$ eV and $V \in (0, 0.6)$ eV.

In Extended Fig. 8, we consider $V = 0.55$ eV and the system of length $L = 200c$ with open boundary conditions. In this case, we find the size of the bulk gaps as $\Delta_{CDW} = 0.53$ eV. For a large range of $\phi$, we observe two edge states in each bulk gap (red dots in Extended Fig. 8a). One is localized at the left end, while the other one is localized at the opposite end. In Extended Fig. 8b, we plot the wavefunction profile of one edge state. Clearly, the wavefunction are exponentially localized at an open boundary. By fitting the wavefunction with an exponential function, we extract the localization length as $r_0 \approx 1.29$ nm. This value is consistent with the one observed at $T = 160$ K. Similarly, if we consider a smaller potential $V = 0.3$ eV, we obtain smaller bulk gaps $\Delta_{CDW} \approx 0.298$ eV and a longer $r_0 \approx 2.64$ nm. These values also agree with the observations at $T = 250$ K. In Extended Fig. 8c, we compute $\Delta_{CDW}$ and $r_0$ for varying $V$, and plot $r_0$ as a function of $\Delta_{CDW}$. We see that $r_0$ is approximately inversely proportional to $\Delta_{CDW}$, which confirms the relationship $r_0 \propto \hbar v_F/\Delta_{CDW}$. Note that near the band center the electron velocity $v_F$ (in the absence of charge density wave) can be taken as a constant. This relationship $r_0 \propto \hbar v_F/\Delta_{CDW}$ also implies that the edge states are protected by the charge density wave gap. In Extended Fig. 8c, we also perform the same calculations for different charge density wave wavelengths, namely, $\lambda = 10c, 15c$ and $20c$. It is clear to observe that $r_0$ is almost independent of $\lambda$. It may also be worth noting that $r_0$ can be very different from $\lambda$. This again indicates that the large charge density wavelength $\lambda$ is irrelevant to $r_0$.

**Reproducibility of our scanning tunneling microscopy data**
Here to test the reproducibility of our data, we used a different tip (commercial Pt/Ir tip) and a different cleaved Ta$_2$Se$_8$I sample for scanning tunneling microscopy and spectroscopy measurements. Extended Fig. 12 shows such data. All the salient features discussed in the main text, namely charge density wave (with a period of $\simeq 12$ nm and $\Delta_{CDW} \simeq 0.55$ eV) and an in-gap edge state within the charge density wave gap are reproduced, as clearly seen in Extended Fig. 12.

**High spatial resolution, large-scale spectroscopic maps revealing d$I$/d$V$ oscillations along the edge**
Here, we examine a long, monolayer step edge and perform detailed spectroscopic mapping as summarized in Extended Fig. 13. Extended Fig. 13a shows the topography and the corresponding d$I$/d$V$ maps taken at different biases. Notably, the line profiles along the edge oscillates with a $\lambda_{avg} = 17.2 \pm 0.4$ nm, i.e., $(1.43 \pm 0.03) \lambda_{CDW}$ periodicity (Extended Fig. 13b). Consistent with Fig. 4's findings, this periodicity is in excellent agreement with the charge density wave period projected along the edge, $\sqrt{2} \lambda_{CDW}$. This recurring d$I$/d$V$ periodicity of the edge state underscores its strong connection with the charge density wave phase. A closer examination of the line profiles reveals that the d$I$/d$V$ oscillation peaks shift along the edge as bias changes. Extended Fig. 13c, which displays the peak positions at different biases, reveals a $4.1 \pm 0.7$ nm, equivalent to a $(0.24 \pm 0.04) \sqrt{2} \lambda_{CDW}$ shift between the -0.2 V and 0.2 V data. The periodicity and the phase shift as a function of bias voltage of the d$I$/d$V$ oscillations can also be seen in the momentum space data presented in Extended Figs. 13d, e which we obtain by Fourier transforming the d$I$/d$V$ linecuts in Extended Fig. 13b. Taken together, the d$I$/d$V$ oscillations with periodicity that closely matches $\sqrt{2} \lambda_{CDW}$ and the shift of the peaks as a function of bias are consistent with Fig. 4 data, and they are consistent with the unusual spectral function proposed in ref. [58]. These findings point to the presence of an exotic edge state that is strongly linked to the charge density wave in Ta$_2$Se$_8$I and its unique topological connectivity.

**Arguments for the topological nature of the edge state**
While examining the possible topological nature of the edge state in Ta$_2$Se$_8$I, we thoroughly considered several possible trivial origins, including (i) an abrupt phase shift of the CDW modulation, (ii) a local potential change, and (iii) a reduction in the CDW strength across the step edge. However, none of these scenarios can account for the intriguing features observed in the experimental data, which include: (1) the edge state's exponential localization at the boundary, (2) its localization length being determined by the CDW gap, and (3) the π phase shift of the edge state when transitioning from the conduction band edge to the valence band edge. Scenario (i) can be confidently ruled out since we did not experimentally observe a phase shift in the CDW modulation across the step edge. Scenarios (ii) and (iii) could potentially trap electrons at the edge with energies within the bulk gap. However, any resulting edge state would significantly depend on local potential changes and reductions in the CDW strength. Consequently, its localization length scale would typically not show a systematic dependence on the bulk gap. Moreover, it is improbable for a trivial edge state to display a CDW modulation along the edge direction, and even more unlikely for the phase of this trivial edge state to transition from the conduction to valence bands. In contrast, as we have discussed, all these unique features are well explained by our minimal effective model. Moreover, we have observed edge states with the same properties in different samples and at different edges. Based on these compelling facts, we firmly believe that the observed edge state has a topological origin that can be effectively captured by our tight-binding model.

To obtain the dispersion of the edge state, we traced the peaks of the edge state at different energies, as shown in Fig. 4. We can see that as the energy (i.e., bias) decreases, the peak position moves in one direction. This clearly shows the spectral flow of edge state along the edge. Furthermore, we find that the extracted phase of the edge state advances by π when the energy goes from the conduction to the valence band edge. These observations are also in good agreement with our theoretical simulations, see Extended Fig. 14. Finally, we would like to point out that while we use the minimal effective model to elucidate the essential physics of the charge density wave phase, the qualitative results are not restricted to this specific model with given topological invariants. In fact, we can refine the model, for example, by enforcing nonsymmorphic symmetry, as discussed in the supplemental material of ref.[58], and show that all the main features persist, but with a different Chern number ($=\pm 1$). While the real samples might be complicated and we cannot completely exclude other (unknown) possible origins of the edge state, our minimal model provides a novel and (so far only) feasible explanation.

**Discussion on the origin of the charge density wave state in $Ta_2Se_8I$**
In this section we provide a discussion on the origin of the charge density wave state in $Ta_2Se_8I$. Existing theoretical and experimental studies considered different scenarios of the charge order formation, encompassing electronic instability[14,33,37], electron-phonon coupling[18,29,62], and a non-Peierls lattice-driven mechanism[21]. However, a comprehensive analysis linking the theoretical developments with the experimental results is required to elucidate the mechanism of the charge density wave transition in $Ta_2Se_8I$. Yet, conducting a full theoretical investigation would require *ab initio* driven theoretical models, which is extremely challenging and cumbersome due to the large unit cell size after incorporating the charge density wave.

In our tunneling experiments we observe a large charge density wave gap ($\Delta_{CDW} \simeq 550$ meV, $\Delta_{CDW}/3.52k_BT_{CDW} \simeq 7$), signaling a strong coupling nature of the charge density wave[35,36]. Neutron scattering[29] and angle-resolved photoemission spectroscopy[62] experiments also provide evidence supporting strong electron-phonon coupling in $Ta_2Se_8I$. Furthermore, judging from the band structures obtained from first principles calculation[37] for bulk $Ta_2Se_8I$, there is no obvious Fermi surface nesting scenario that can match the experimental observations. Fermi surface nesting (if feasible) based on the quasi-one-dimensional band structure would indicate that the charge density vector is aligned with the *c*-axis and has a magnitude of about $\pi/c$. However, in the tunneling[20,21] and diffraction experiments[23-29], the charge density wavevector is observed to have all three coordinate components as finite and directed in the *ab-c* direction with its magnitude being very small. We also note that the spin-orbit coupling in this material is rather small, with an energy scale on the order of 20 meV[14,37,58]. Consequently, the relevant energy window for the Weyl semimetal phase (when the small spin-orbit coupling is taken into account) is also very narrow[14]. Thus, the energy scale associated with Weyl physics appears to be significantly smaller than the observed charge density wave gap. Therefore, it is unlikely that the nesting between Weyl points serves as the driving force behind the charge density wave in $Ta_2Se_8I$. Conversely, since no direct evidence refutes electron-phonon coupling as the primary force behind the transition, it remains a plausible mechanism for the charge density wave transition. Therefore, a comprehensive understanding of the charge density wave's origin in this material requires further investigation, which is beyond the scope of this manuscript.

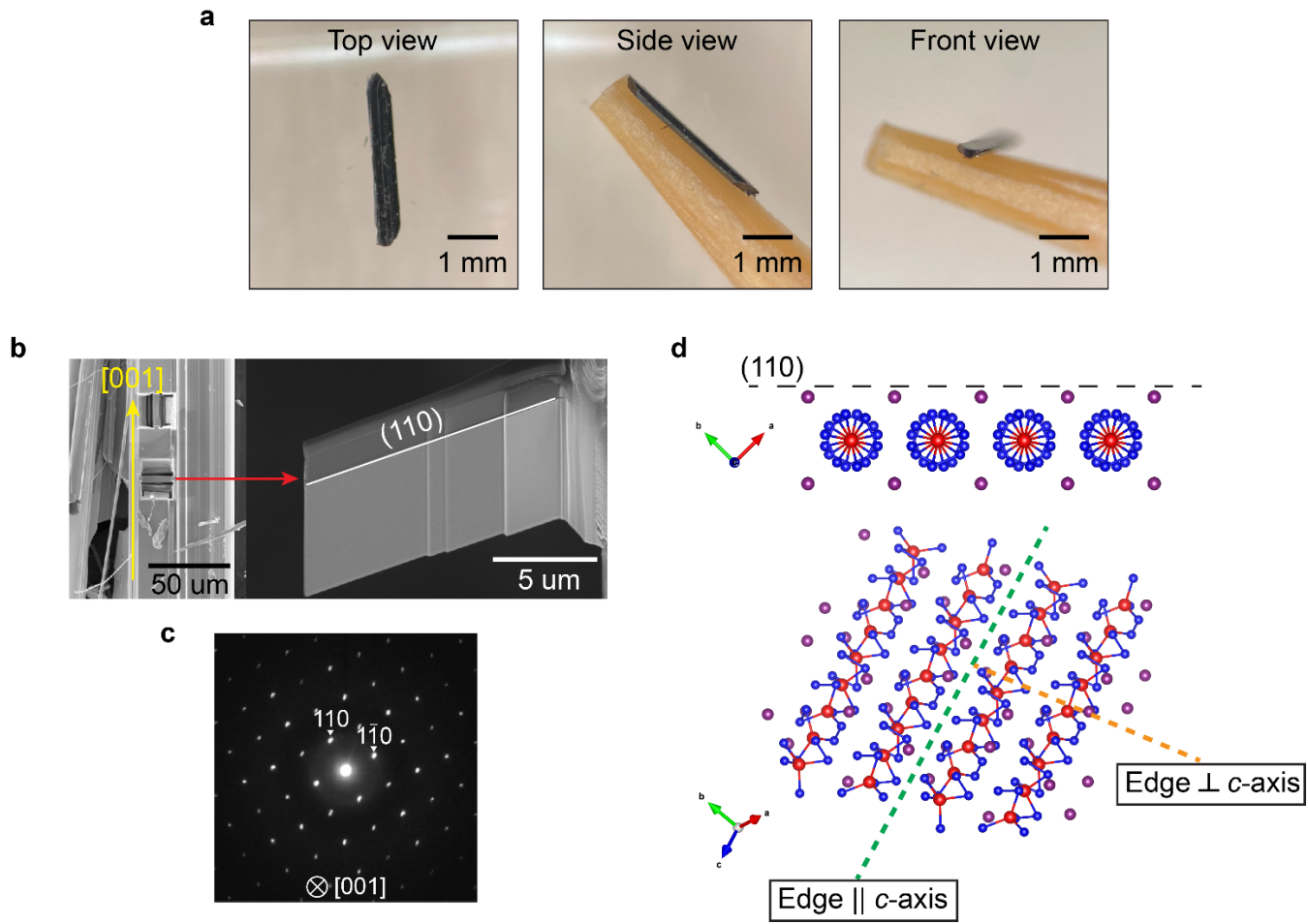

**Extended Fig. 1: Ta$_2$Se$_8$I cleavage plane identification. a,** Optical microscope images of a typical Ta$_2$Se$_8$I crystal showing top, side, and front view. Crystals have a geometrical shape of a thin long bar, where only two equivalent, large area facets are accessible for cleavage. **b,** Scanning electron microscopy image of a thin lamellae prepared by focused ion beam cutting from the cleavage surface of bulk Ta$_2$Se$_8$I. **c,** Corresponding electron diffraction pattern indicating that the cleavage plane is (110). **d,** Top: *ab* plane projection of Ta$_2$Se$_8$I crystal structure. (110) cleavage plane is marked with black dashed line. Bottom: 3D view of the crystal structure, showing two potential edge orientations. Step edges along the *c*-axis (edge direction is marked with dashed green line) are favored as no covalent bond breaking is required, while step edges perpendicular to the *c*-axis (edge direction is marked with dashed orange line) are *not* energetically favored as they would require breaking many strong covalent atomic bonds.

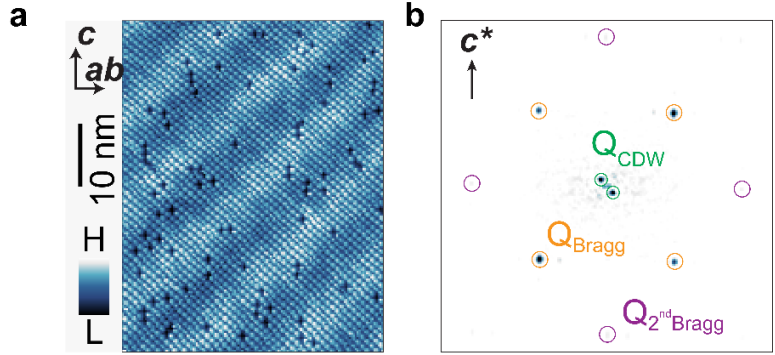

**Extended Fig. 2: Scanning tunneling microscopy topographic imaging of charge density wave in Ta$_2$Se$_8$I. a,** Atomically resolved, large-scale topographic image of the (110) plane exhibiting a clear charge density wave ($V_{gap}$ = -1.0 V, $I_t$ = 50 pA). **b,** Fast Fourier transform of the topographic image capturing the first (orange circles) and second (purple circles) order Bragg peaks as well as pronounced charge density wave peaks (marked with green circles).

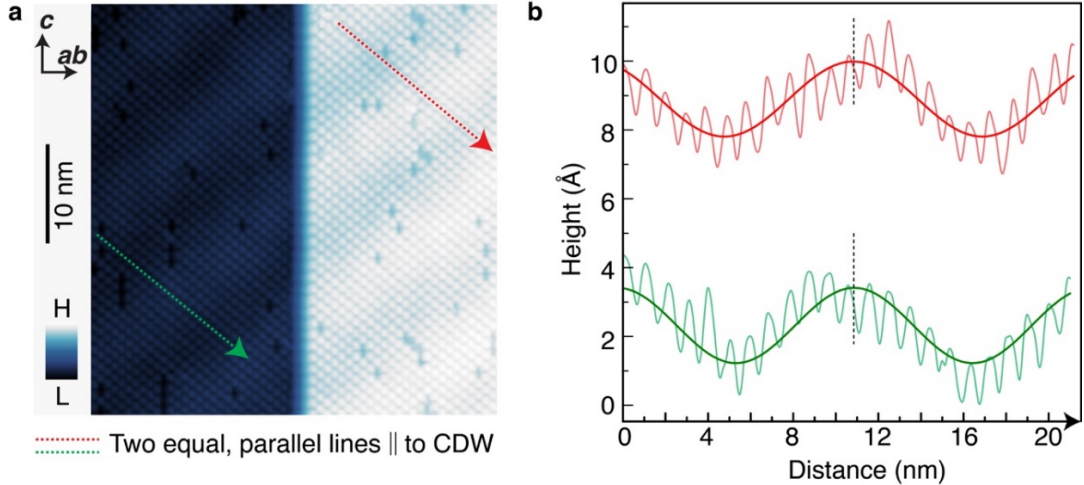

**Extended Fig. 3: Evolution of the charge density wave phase as it encounters a monolayer step edge. a,** Same topography as in Fig. 2a showing an atomically sharp monolayer step edge. Two color-coded (red and green), parallel lines of equal length are drawn along the charge density wave direction and on the two sides of the monolayer step edge. **b,** The color-coded line profiles taken along the red and green lines (the direction of the scan is marked with arrows), exhibit clear, uniform atomic and charge density wave correlations. To capture the charge density wave envelope superimposed on the atomic features, we perform a fitting to sinusoid (shown using color-coded dashed curves). Their maxima positions are marked with dashed vertical lines. There is no detectable phase shift between the two sinusoids, implying a phase continuity of the charge density wave as it crosses the monolayer step edge.

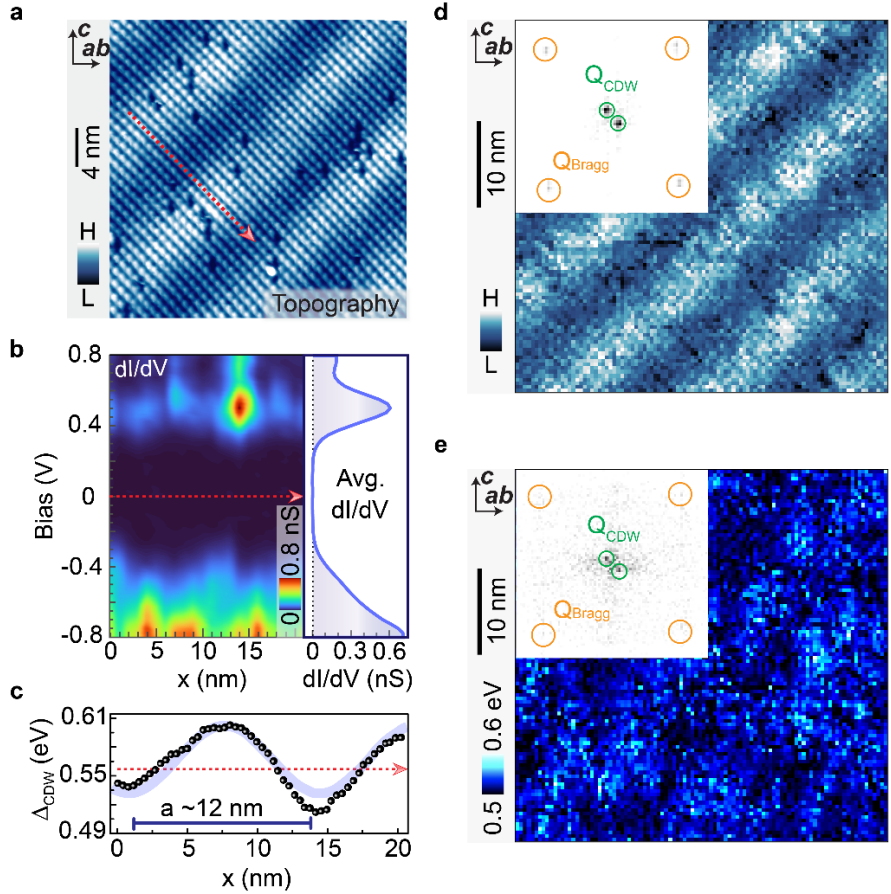

**Extended Fig. 4: Spatially resolved differential spectra displaying a periodic energy gap modulation closely mirroring the periodicity of the charge density wave. a**, Atomically resolved topographic image of the (110) plane exhibiting a clear charge density wave with $\lambda_{CDW} \simeq 12\ nm$ ($V_{gap}$ = -1.0 V, $I_t$ = 50 pA). **b**, Intensity plot of a series of line spectra taken along the charge density wave direction (marked on the corresponding topographic image in panel **a** with a red line; the direction of the scan is marked with an arrow). The averaged differential spectrum, shown on the right, reveals a large insulating gap (the dotted vertical line denotes zero differential conductance). Tunneling junction set-up: $V_{set}$ = -0.8 V, $I_{set}$ = 0.4 nA, $V_{mod}$ = 10 mV. **c**, Energy gap values extracted from the series of line spectra in panel **b** revealing modulation of the energy gap as a function of distance along the charge density wave direction. The pale purple curve is the fitting of $\Delta_{CDW}(x)$ with $A \sin \frac{2\pi}{\lambda}(x - \phi)$. Extracted parameter $\lambda \simeq 12\ nm$ is the gap modulation period. This modulation period matches with the charge density wave periodicity obtained from the topographic image, thus showing consistency with an electronic nature of the charge density wave. **d** and **e**, Topography and the corresponding spatially resolved energy gap map, revealing charge density wave modulation. Insets show the Fourier transforms of the respective data. The data in panels **a-e** are taken at $T$ = 160 K.

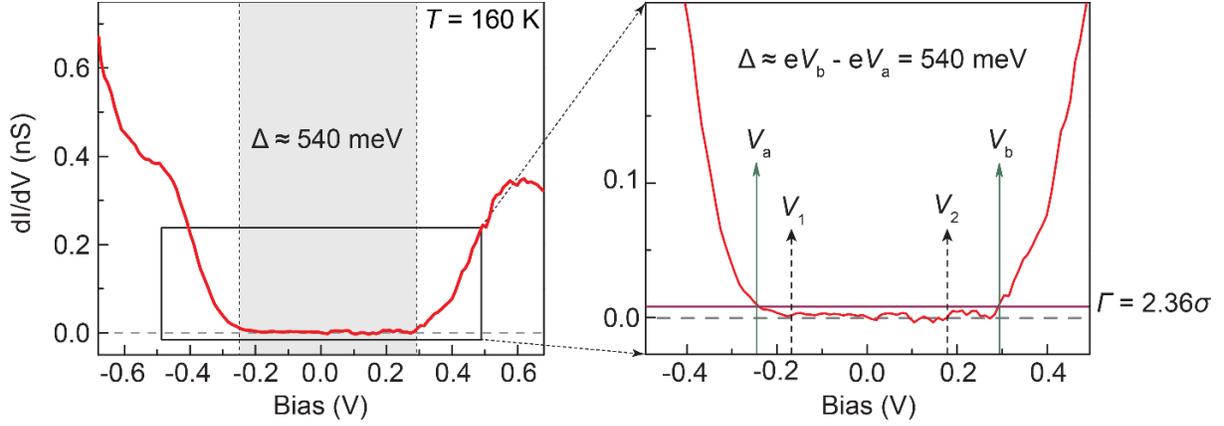

**Extended Fig. 5: Determination of the energy gap from d$I$/d$V$ spectrum.** Left: Representative d$I$/d$V$ spectrum acquired on Ta$_2$Se$_8$I (110) surface. Right: Magnified view of the gapped region in the d$I$/d$V$ spectrum shown on the right. $[V_1, V_2]$ is the voltage interval used for the calculation of the noise floor, $\sigma$. $\Gamma = 2.36\sigma$ (purple line) is set as the instrumental resolution of the d$I$/d$V$ signal. $V_a$, $V_b$ are numeric solutions of the equation d$I$/d$V = \Gamma$. A spectroscopic energy gap is determined as $\Delta = eV_b - eV_a$, which is approximately 540 meV in this case.

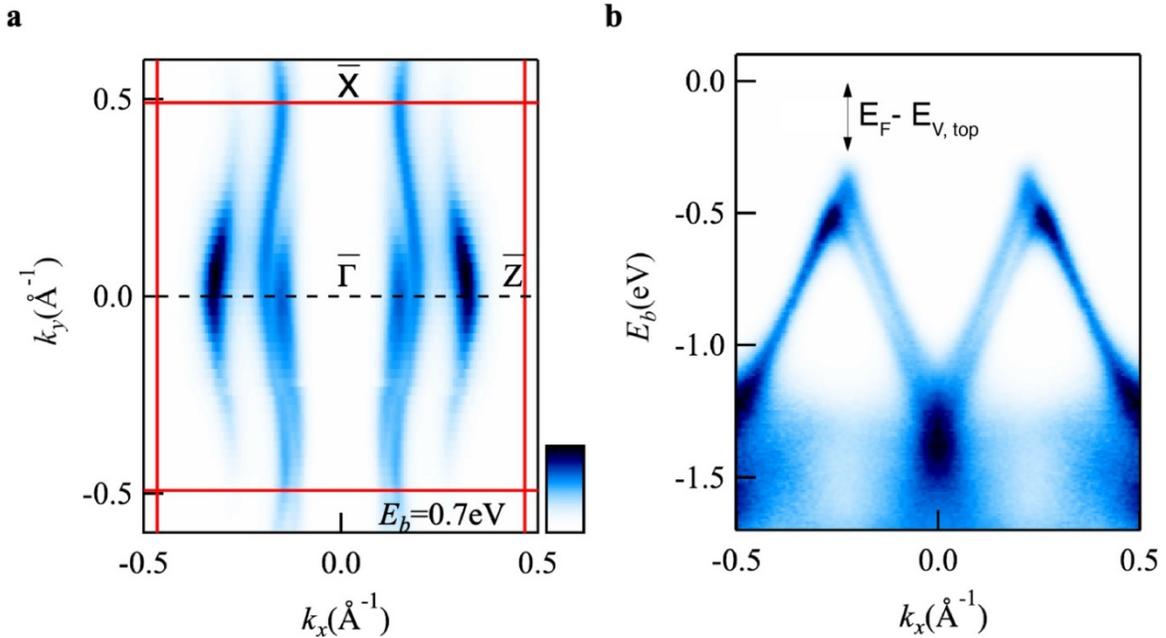

**Extended Fig. 6: Band structure of Ta$_2$Se$_8$I in the charge density wave phase obtained from photoemission spectroscopy**. **a**, Constant energy contour at binding energy $E_b$ = 0.7 eV. **b**, Energy momentum cut along the $\bar{\Gamma} - \bar{Z}$ direction (shown using a dashed black line in **a**), where the lower branch of the bulk Dirac cones can be directly observed. The band splitting occurs due to the surface charge effect; this phenomenon has been discussed in ref. [20]. The spectra are acquired using a right circularly polarized 30 eV light at 85 K. We symmetrized the spectra with respect to $k_x = 0$.

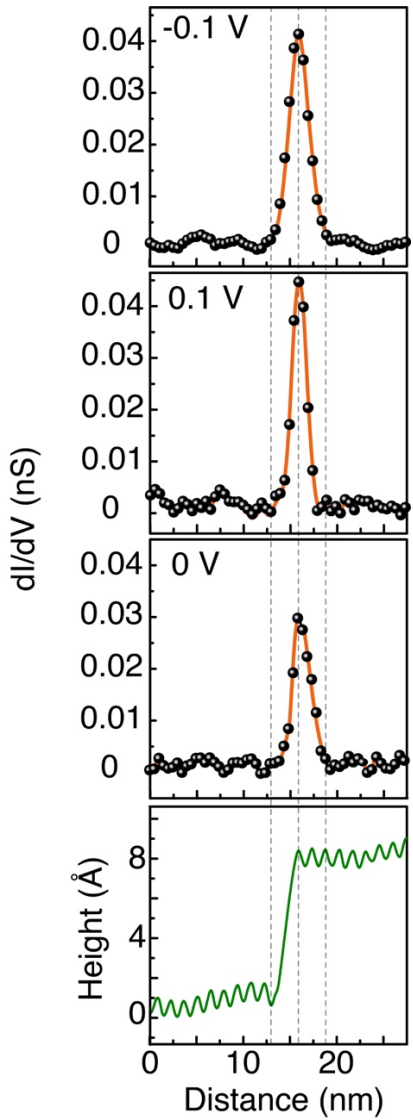

**Extended Fig. 7: Line profiles of topography and dI/dV around a monolayer atomic step edge.** Bottom panel: The height profile obtained from a topographic image shown in Fig. 2**a**, taken perpendicular to the *c*-axis direction, exhibits uniform atomic correlation perpendicular to the step edge over a large distance. Top panels: Intensity distribution of d*I*/d*V* taken at the same locations within the insulating gap ($V$ = 0 mV and ±100 mV) reveal a pronounced edge state. The intensity distributions exhibit maxima at the edge, exponential decay away from the step edge on the crystal side, and a sharp decay on the vacuum side, falling to near zero value right where the height profile drops to its minimum.

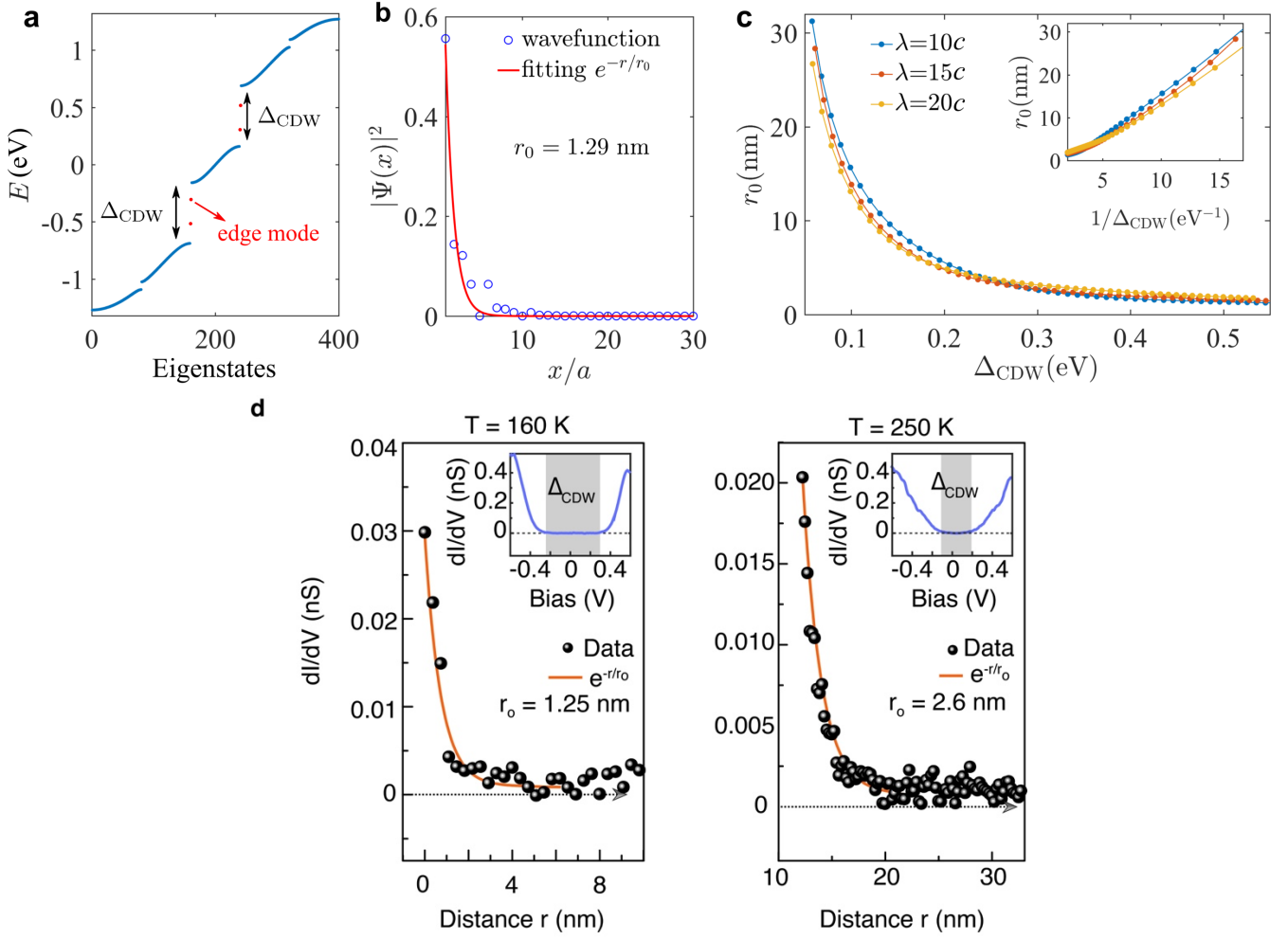

**Extended Fig. 8: Localization length of edge state. a,** Energy spectrum of the charge density wave chain with open boundary conditions. The red dots represent the edge states. **b,** Wavefunction profile of the edge state at $E \approx -0.32$ eV. The edge state is exponentially localized at one open boundary. The localization length is extracted as $r_0 \approx 1.29$ nm. **c,** $r_0$ as a function of $\Delta_{CDW}$ for $\lambda = 10c$, $15c$ and $20c$, respectively. Inset: $r_0$ as a function of $1/\Delta_{CDW}$. We consider $V = 0.55$ eV, $\lambda = 10c$ in **a**, $c = 1.3$ nm, $t = 1.2$ eV and $\phi = 0.5\pi$ in all panels. **d,** Experimental d$I$/d$V$ line profile perpendicular to the edge showing the decay length at $T = 160$ K ($\Delta_{CDW} \simeq 0.55$ eV) and $T = 250$ K ($\Delta_{CDW} \simeq 0.3$ eV), respectively. Insets show the corresponding d$I$/d$V$ spectra acquired on the surface.

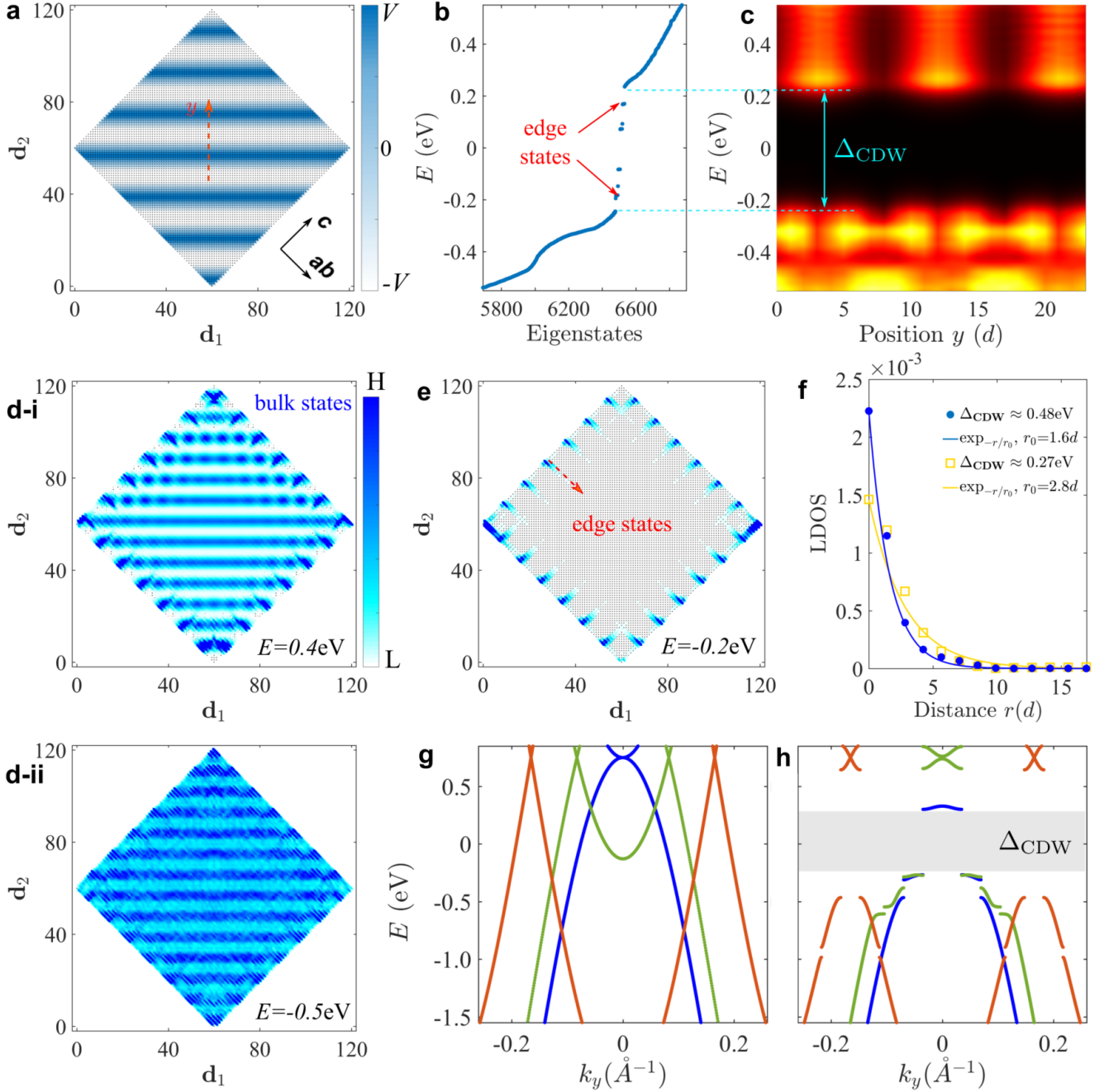

**Extended Fig. 9: Topological charge density wave phase of the theoretical model. a,** Schematic of the theoretical model with the color indicating the variation of the charge density wave potential. **b,** Low-energy spectrum of the system. The charge density wave potential generates bulk gaps of size around $\Delta_{CDW} \simeq 0.48$ eV, inside which in-gap edge states emerge. **c,** LDOS along a line inside the bulk (indicated by the red dashed arrow line in **a**). **d-i,** LDOS at the energy $E = 0.4$ eV (measured from $E_F$) in the conduction continuum. **d-ii,** The same as **d-i** but for the energy $E = -0.5$ eV in the valence continuum. The LDOS minima (maxima) at the conduction band correspond to the LDOS maxima (minima) at the valence band, signifying a contrast reversal in the charge density wave pattern. **e,** LDOS at the energy $E = -0.2$ eV in the bulk gap. **f,** LDOS at $E = -0.2$ eV as a function of position $r$ along the line from an edge into the bulk (indicated by the red dashed arrow line in **e**). The density profile decays exponentially from the edge. The blue solid curve is an exponential fitting with a localization length $r_0 =$

$1.6d$. The yellow squares and solid curve are the same as the blue ones but for a smaller gap $\Delta_{CDW} \simeq 0.27$ eV (corresponding to $V = 0.6$ eV). The yellow curve represents an exponential fitting with a localization length $r_0 = 2.8d$. **g**, The energy dispersions against $k_y$ in the absence of $V$ for $k_x = 0$ (blue), $\pi/4a$ (green) and $\pi/2a$ (orange), respectively. All sub-bands (characterized by different $k_x$) develop substantial gaps around the Fermi energy in the presence of the charge density wave potential. Thus, the system has a full charge density wave gap. **h**, The same as **g** but with a finite charge density wave potential $V = 1.2$ eV. The gray area indicates the charge density wave insulating gap in the system. Other parameters are $\phi = 0.2\pi$, $a = 0.95$ nm, $\lambda = 18d$, $T = 200$ K, $t_x = t_y = 1.5$ eV, $V = 1.2$ eV and $E_F = -0.7$ eV.

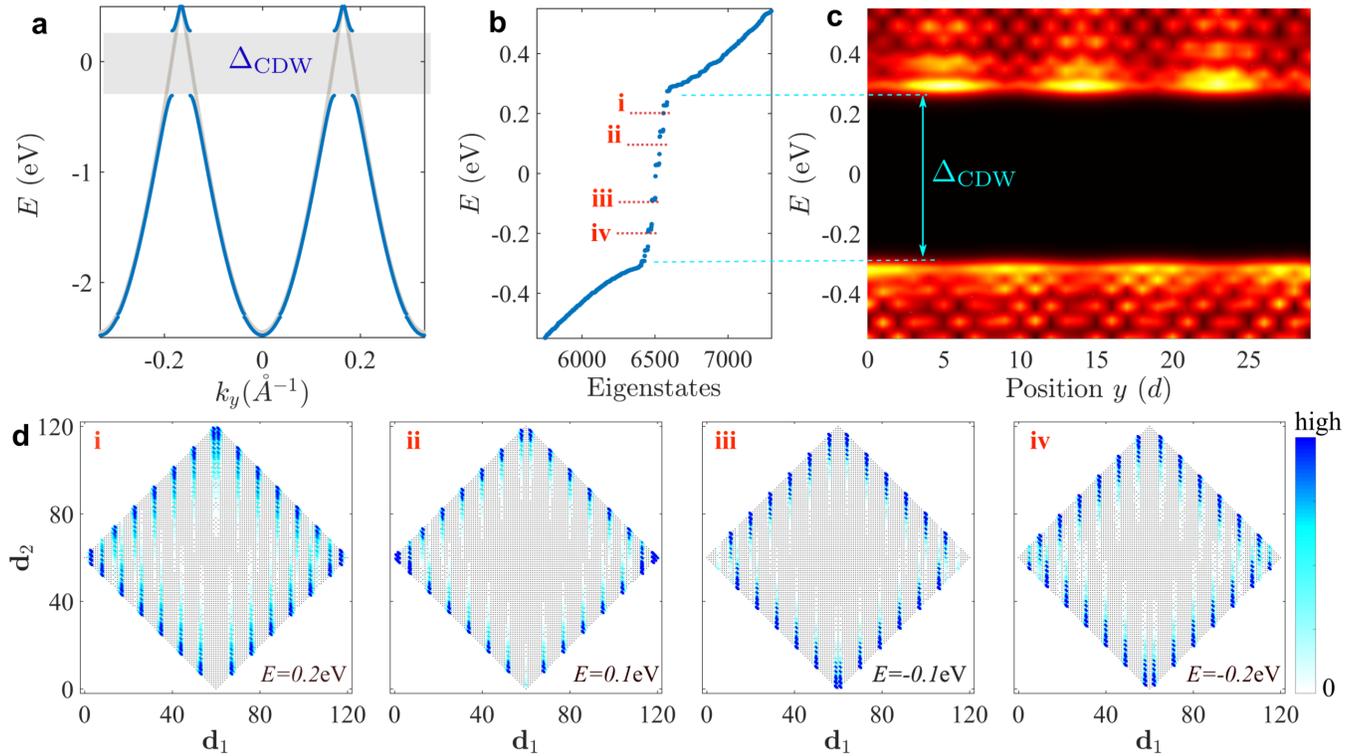

**Extended Fig. 10: Edge states within the charge density wave gap. a,** The energy dispersion against $k_y$ in the decoupled wire limit ($t_x = 0$). The gray curves are the dispersion in the absence of the charge density wave. The gray area indicates the insulating gap induced by the charge density wave in the system. **b-c,** The same as Extended Figs. 9**b, c** but for the decoupled wire limit. **d**, LDOS at four in-gap energies $E = -0.2$ eV, $-0.1$ eV, $0.1$ eV and $0.2$ eV, respectively. $V = 0.6$ eV, $E_F = -0.55$ eV and other parameters are the same as Extended Fig. 9.

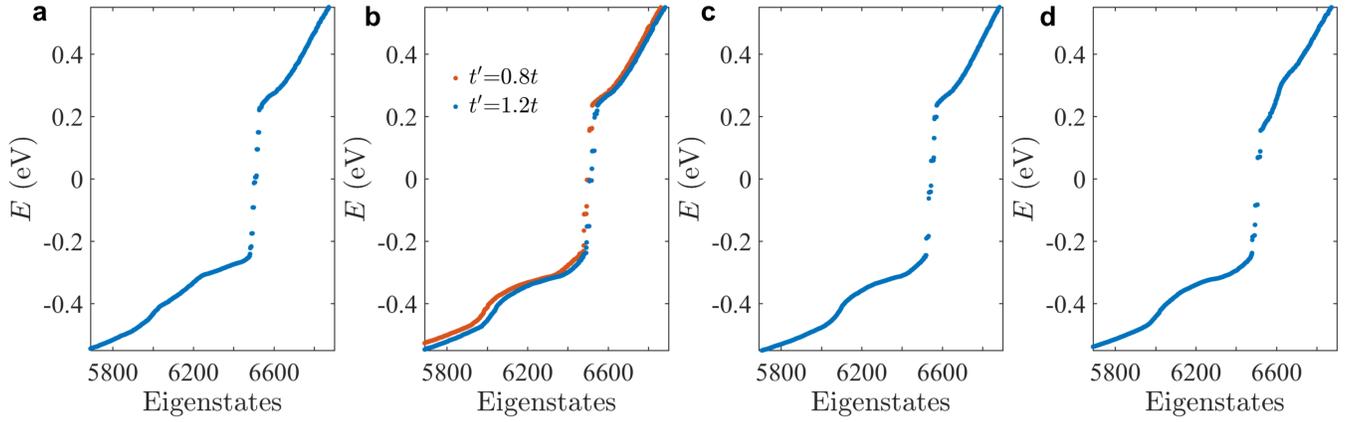

**Extended Fig. 11: Low-energy spectra of the system with additional perturbation effects**. **a,** Low-energy spectrum in the presence of a small net average charge density wave potential $V_0 = 0.1V$ (i.e., $V(\mathbf{r})$ varies from $-V + V_0$ to $V + V_0$ when moving in space). **b,** Low-energy spectrum in the presence of enhancement ($t' = 1.2t$) or reduction ($t' = 0.8t$) of nearest-neighbor hopping associated with lattices that are closest to and second closest to the boundary due to lattice reconstruction close to the boundary. **c,** Low-energy spectrum in the presence of a local potential $t\sigma_z$ for lattices that are closest to and second closest to the boundary. **d,** Low-energy spectrum in the presence of a small next-nearest-neighbor hopping $t_2 = 0.02t$. Other parameters are the same as Extended Fig 9.

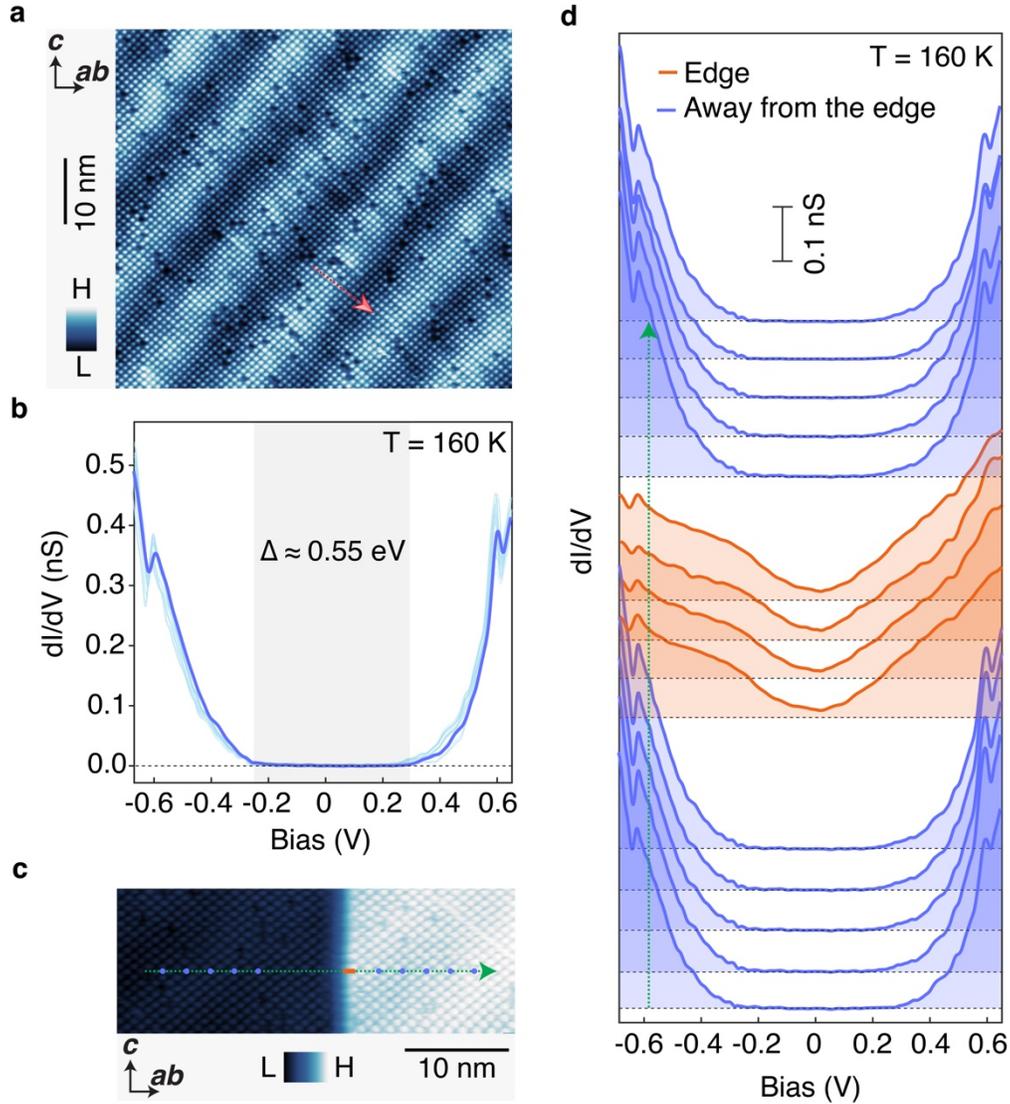

**Extended Fig. 12: Reproducibility of the experimental data – Topography and spectroscopy data in a separate Ta$_2$Se$_8$I sample, acquired using a commercial Pt/Ir tip. a**, Atomically resolved topographic image of the (110) plane exhibiting a clear charge density wave ($V_{gap}$ = -1.0 V, $I_t$ = 50 pA). **b**, A series of differential spectra (shown using light blue curves) taken along the charge density wave direction (marked on the corresponding topographic image in panel **a** with a red line; the direction of the scan is marked with an arrow) Tunneling junction set-up: $V_{set}$ = -0.8 V, $I_{set}$ = 0.4 nA, $V_{mod}$ = 10 mV. The averaged differential spectrum (dark blue curve) reveals a large insulating gap, consistent with Fig. 1**k** of the main text. The dotted horizontal line denotes zero differential conductance. **c**, Atomically resolved topography of an atomically sharp monolayer step edge ($V_{gap}$ = -1.0 V, $I_t$ = 50 pA). **d**, d$I$/d$V$ spectra revealing an insulating gap away from the edge and a pronounced in-gap state on the edge. Tunneling junction set-up: $V_{set}$ = -0.8 V, $I_{set}$ = 0.4 nA, $V_{mod}$ = 10 mV. Orange and blue curves represent the differential spectra taken at different positions (marked on the panel **c**) at the edge and away from the edge, respectively. Spectra are offset for clarity. Dotted horizontal lines denote zero differential conductance. All the data are taken at $T$ = 160 K.

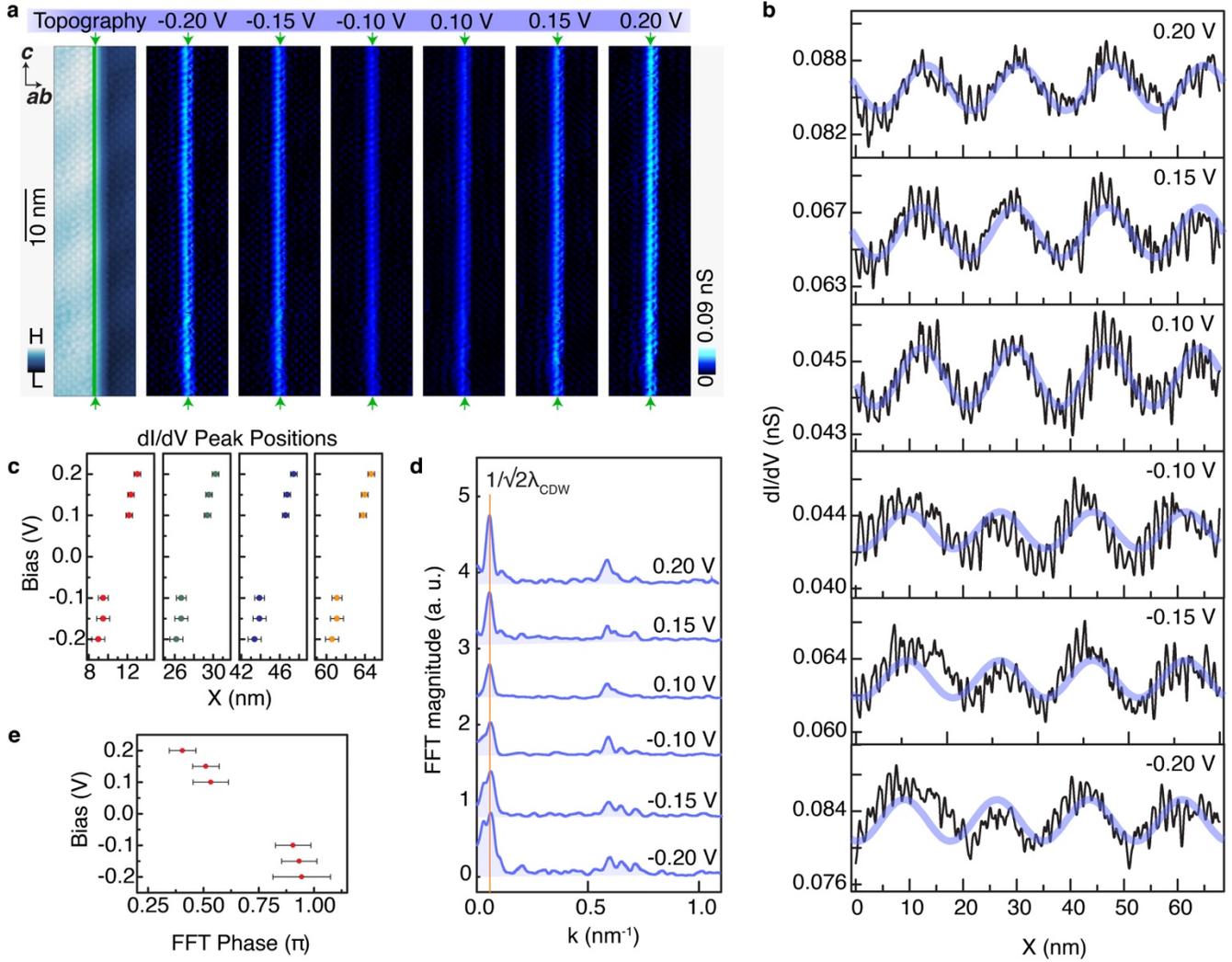

**Extended Fig. 13: High spatial resolution, large scale spectroscopic maps, visualizing periodic oscillation of differential spectra along the edge. a,** Atomically resolved topography and the corresponding differential conductance maps around a monolayer step edge taken at six different bias voltages showing the presence of an edge state. The green arrows mark the location of the edge along the ***ab*** axis. Tunneling junction set-up for d$I$/d$V$ maps: $V_{set}$ = -0.8 V, $I_{set}$ = 0.3 nA, $V_{mod}$ = 10 mV. All data are taken at $T$ =160 K. The charge density wave in this specific sample displays a periodicity of $\lambda_{CDW}$ = 12.0 ± 0.6 nm, which has been deduced from topographic imaging of a region located away from the edge. **b,** d$I$/d$V$ line profiles (averaged over an atom along the edge marked with a green vertical line in the topographic image in panel **a**) along the edge at six biases ranging from -0.2 V to 0.2 V exhibiting a periodic modulation which can be fitted with $A \sin\frac{2\pi}{\lambda}(x - \phi)$; fitting parameters are $\lambda$ and $\phi$ which denote wavelength and phase, respectively. Averaging $\lambda$ obtained from the fitted sinusoids of the six curves, we obtain $\lambda_{avg}$ = 17.2 ± 0.4 nm. The pale purple curves denote fitting of $A \sin\frac{2\pi}{\lambda_{avg}}(x - \phi)$ applied to d$I$/d$V$ data at each bias. **c,** Peak positions of the d$I$/d$V$ oscillations (obtained from the fit marked with pale purple curves) plotted for different biases. The locations of the peaks shift as a function of bias. Quantitatively, there is a 4.1 ± 0.7 nm shift between the profiles at -0.2 V and 0.2 V. **d and e,** Momentum space representation of the line profile data. **d,** Fourier transform magnitude of the line profiles exhibits a peak representative of the periodicity of

the d$I$/d$V$ oscillations along the edge. The vertical orange line marks the location of the charge density wave spatial frequency projected along the edge, i.e., $1/\sqrt{2}\,\lambda_{CDW}$. **e,** Fourier transform phase, obtained at the Fourier transform magnitude peaks at different bias voltages, shows a clear phase shift as a function of bias.

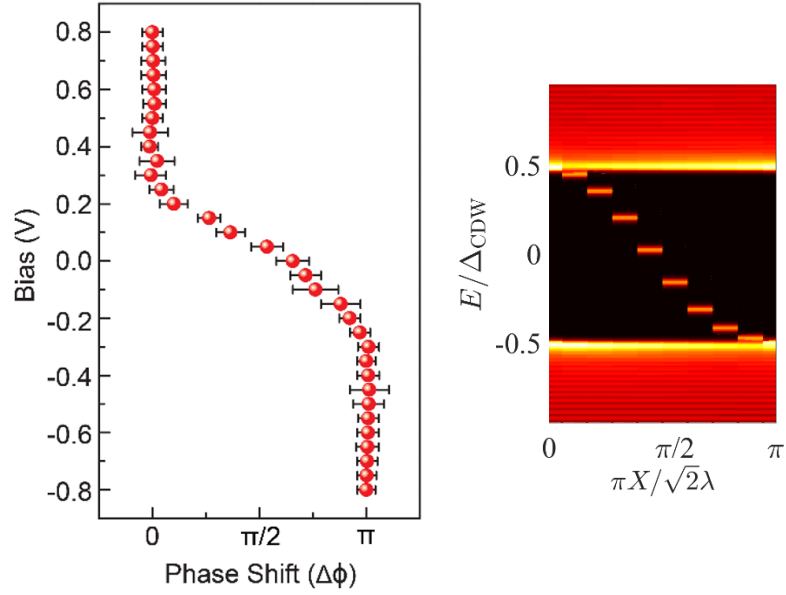

**Extended Fig. 14: Comparative analysis of theoretical and experimental phase shift at the edge as a function of energy.** Left: Deduced phase shift $\Delta\phi = \phi - \phi_{V=0.8V}$ of the d$I$/d$V$ oscillations in the experiment, adapted from Fig. 4**c**. Right: Theoretically calculated local density of states along an edge direction $X$ in a ribbon geometry with parameters $\lambda' = 2\lambda = 18d$ and $V = 0.2t$.

**Competing interests:** The authors declare no competing interests.

**Data and materials availability:** All data needed to evaluate the conclusions in the paper are present in the paper. Additional data are available from the corresponding authors upon reasonable request.

**Acknowledgement:** Experimental and theoretical work at Princeton University is supported by Gordon and Betty Moore Foundation (GBMF4547 and GBMF9461). Crystal growth at Max Planck Institute for Chemical Physics of Solids is funded by European Research Council (ERC) Advanced Grant No. 742068 ("TOPMAT"), Deutsche Forschungsgemeinschaft (DFG) under SFB 1143 (Project No. 247310070), and Würzburg-Dresden Cluster of Excellence on Complexity and Topology in Quantum Matter—ct.qmat (EXC 2147, Project No. 39085490). Crystal growth at Beijing Institute of Technology is supported by National Science Foundation of China (NSFC) (11734003) and the National Key Research and Development Program of China (2016YFA0300600). T.A.C. was supported by the National Science Foundation Graduate Research Fellowship Program under grant no. DGE-1656466. Y.G.Y. is supported by NSFC (11574029) and the Strategic Priority Research Program of Chinese Academy of Sciences (XDB30000000). T.N. acknowledges supports from the European Union's Horizon 2020 research and innovation programme (ERC-StG-Neupert-757867-PARATOP).